\documentclass[10.5pt,preprint]{aastex}
\usepackage{apjfonts}

\usepackage{epsfig} 

\newcommand\etal{{et~al.}} 

\slugcomment{Accepted for publication in ApJ}
\shorttitle{Nonlinear Absorption-line Index Versus Metallicity Relations of M31 Globular Clusters}
\shortauthors{Kim et al.}

\begin{document}

\title{Nonlinear Color--Metallicity Relations of Globular Clusters. V. Nonlinear Absorption-line Index versus Metallicity Relations and Bimodal Index Distributions of M31 Globular Clusters}

\author{Sooyoung Kim\altaffilmark{1}, Suk-Jin Yoon\altaffilmark{1}, Chul Chung\altaffilmark{1}, Nelson Caldwell\altaffilmark{2}, Ricardo P. Schiavon\altaffilmark{3}, Yongbeom Kang\altaffilmark{4}, Soo-Chang Rey\altaffilmark{4}, and Young-Wook Lee\altaffilmark{1}}

\altaffiltext{1}{Department of Astronomy and Center for Galaxy Evolution Research, Yonsei University, Seoul 120-749, Republic of Korea}
\email{sjyoon@galaxy.yonsei.ac.kr}
\altaffiltext{2}{Center for Astrophysics, 60 Garden Street, Cambridge, MA 02138}
\altaffiltext{3}{Gemini Observatory, 670 North A'ohoku Place, Hilo, HI 96720} 
\altaffiltext{4}{Department of Astronomy and Space Science, Chungnam University, Daejeon, Republic of Korea} 

\begin{abstract}
Recent spectroscopy on the globular cluster (GC) system of M31 
with unprecedented precision witnessed a clear
bimodality in absorption-line index distributions of old GCs.
Such division of extragalactic GCs, so far asserted mainly by 
photometric color bimodality, has been viewed as the presence of merely two distinct metallicity subgroups within individual galaxies
and forms a critical backbone of various galaxy formation theories.
Given that spectroscopy is a more detailed probe into stellar
population than photometry, the discovery of index bimodality may
point to the very existence of dual GC populations. However, here
we show that the observed spectroscopic dichotomy of M31 GCs emerges
due to the nonlinear nature of metallicity-to-index conversion
and thus one does not necessarily have to invoke two separate GC
subsystems. 
We take this as a close analogy 
to the recent view that metallicity--color nonlinearity is primarily 
responsible for observed GC color bimodality.
We also demonstrate that 
the metallicity-sensitive magnesium line displays non-negligible metallicity--index nonlinearity
and Balmer lines show rather strong nonlinearity.
This gives rise to bimodal index distributions, 
which are routinely interpreted as bimodal {\it metallicity} distributions, 
not considering metallicity--index nonlinearity.
Our findings give a new insight
into the constitution of M31's GC system, which could change much of
the current thought on the formation of GC systems and their host galaxies.

\end{abstract}


\section{INTRODUCTION}

Systems of globular clusters (GCs) have long served the role of
providing important constraints on theories of galaxy formation and
assembly.  
GCs, which comprise stars with small internal dispersion in age and abundance, 
are always found in large galaxies and are easily
identified out to large galactocentric distances due to their compact
sizes and luminosities.
A long-standing puzzle in the field of
galaxy evolution is that GC systems exhibit bimodal color distributions
\cite[e.g.,][]{geisler96, kundu01, larsen01, peng04, harris06,
peng06, lee08a, jordan09, sinnott10, liu11, blakeslee12}, the origin of which
has been the topic of much interest throughout the last couple of
decades.

Colors of GCs measure their integrated spectral slopes, which are
mainly governed by their ages and metallicities. Because the majority
of GCs in galaxies are old \cite[e.g.,][]{forbes01, brodie05,
strader05, cenarro07, proctor08, dotter11}, it is believed that metallicity plays the dominant role
in governing cluster colors. Cluster metallicities are inferred
from colors based traditionally on linear or mildly curved \citep{harris02, cohen03} conversion between color and metallicity. 
Hence, the color bimodality of GC systems
has been widely interpreted as the manifestation of metallicity
bimodality. The notion of two metallicity groups of GCs within
individual galaxies forms a critical backbone of diverse GC formation
theories in the context of galaxy evolution (see Harris 1991; West
\etal\ 2004; Brodie \& Strader 2006 for reviews). Three widely
accepted hypotheses include mergers, mutiphase formation, and
accretion, all of which attribute the colour bimodality to the
existence of two GC subpopulations with different origins
\cite[e.g.,][]{ashman92, cote98, forbes97, mglee10b}.

In contrast, an alternative scenario has been put forward
that does not necessarily invoke the existence of only two separate GC
metallicity subgroups (Yoon et al. 2006, hereafter Paper I).
Recent observations and modeling of the combined GC data of Milky
Way and Virgo elliptical galaxies revealed appreciable nonlinearity in
their color--metallicity relationship (Peng \etal\ 2006; PaperI; Cantiello \& Blakeslee 2007).
Paper I showed that the nonlinearity is a result of two complementary effects: 
the systematic variation in the temperature distribution of 1) red-giant-branch (RGB) stars and 
2) horizontal-branch (HB) stars which are both nonlinear functions of cluster metallicity. 
Such nonlinear nature can create a bimodal color distribution from
a broad underlying metallicity spread, even if it is unimodal (Paper
I; Cantiello \& Blakeslee 2007). The hypothesis asserts that color
spread is a projected distribution of metallicity and therefore any
feature on color--metallicity relations (CMRs) translates to color histogram.

Because the choice of colors determines the shape of the corresponding
CMRs, Yoon \etal\ (2011a, hereafter Paper II) and Yoon \etal\ (2013, hereafter Paper IV)
explored the varying degrees of nonlinearity in CMRs 
using {\it Hubble Space Telescope} ({\it HST})
multiband photometry of the GC systems in M87 (Paper II) and M84 (Paper IV).  
The multiband study revealed that $u$-band related colors in particular
may provide insight into the underlying cluster metallicity distribution function (MDF).  
They demonstrated that different
combination of $ugz$-bandpasses produces different color histogram
morphologies under the nonlinear-CMR assumption, which is in good agreement
with the observed color distributions of both M87 and M84 GC systems.
Also very interestingly, \cite{chies12} and \cite{blakeslee12} showed
that GC systems characterized by optical color bimodality do not
necessarily display bimodality in optical--NIR colors, suggesting
that the color bimodality may indeed not translate directly into
metallicity bimodality.  The nonlinear-CMR scenario was further
put to the test in the subsequent study of GC metallicities and
their link to the host galaxy (Yoon \etal\ 2011b, hereafter Paper
III). By applying a nonlinear color-to-metallicity conversion,
it was shown that the inferred unimodal metallicity distributions 
of GCs with broad metal-poor tails are remarkably
similar to those of halo field stars. This result sheds new light on a
long standing discrepancy between the observed metallicity distributions of GCs
and halo stars. Despite the wide implications of
the nonlinear-CMR scenario for color bimodality entailed in these
studies (Papers I, II, III and IV), the question of whether true CMR
nonlinearity exists is still an open question in need of confirmation from independent observations.

As the nearest giant external galaxy with a large population of
GCs, the Andromeda galaxy, M31, is an obvious and ideal target for studies of the structure
of extragalactic GC systems. The wealth of photometric data
available on M31 GCs can provide important constraints on the genesis
of that well studied GC system. However, despite its unique and
numerous advantages as a prominent member of the Local Group, M31
possesses substantial internal extinction.  Some of M31 GCs are
thus heavily reddened, with the measured values of $E(B-V)$ as high
as 1.3 mag \citep{barmby00, fan08, caldwell11, kang12}. Consequently,
integrated colors of M31 GC system are susceptible to uncertainties
in internal reddening, which in turn hinders a more accurate
analysis of its CMR.  Spectral indices, on the other hand, suffer little from extinction and thus provide a more reliable probe to
explore the nonlinearity issue.
Large spectroscopic samples of high-quality extragalactic GC data
are hard to come by and acquiring them is usually a time consuming
process.  
Because of the relative proximity of M31, however, 
there are a number of existing catalogs
of spectroscopy for its GC system by various authors including
\citet{perrett02}, \citet{barmby00}, \citet{galleti07,galleti09},
\citet{kim07}, \citet{lee08b}, and \citet{caldwell09, caldwell11}.

We use new, high signal-to-noise (S/N) spectroscopic
data on M31 GC system by \citet{caldwell09, caldwell11}. 
This database consists of a large number of M31 cluster spectra, and it provides improved membership classification
and age estimation. By virtue of their high S/N, Caldwell et
al.'s data reveal a clear dichotomy within the old ($>$\,10 Gyr),
spheroidal component of M31 GCs: This spectroscopic dichotomy
is important, given that the GC division has so far been asserted
mainly by a photometric characteristic of GCs, i.e., their color
bimodality. Spectroscopy offers a by far more detailed probe into
the effect of HB morphology on the integrated properties of stellar
contents in GCs than broadband colors. Balmer absorption lines
(H$\beta$, H$\gamma$, and H$\delta$) in particular, are strong indicators
of hot HB stars due to their great sensitivity to temperature
(Worthey 1994, hereafter W94; Schiavon et al. 2004).

Motivated by the observed spectroscopic dichotomy of the old, spheroidal component of M31 GC system \citep{caldwell11}, 
we explore their distributions of spectral-line indices using nonlinear metallicity-to-index conversions.
The paper is the fifth in the series on nonlinear color--metallicity relations 
of extragalactic GC systems and organized as follows. Section 2 describes the data
used in the study. Section 3 makes a brief revisit to the nonlinear-CMR theory 
for color bimodality and explores nonlinearity in the index--metallicity plane.
In Sections 4 and 5 we elucidate the spectroscopic division of GCs
in M31. Section 4 shows how the presence of HB stars affect the
strengths of absorption line indices, causing notable nonlinearity
in the index--index planes. Section 5 carries out the nonlinear
conversion from metallicity to line indices using (1) an assumed
single gaussian distribution for a metallicity distribution [Fe/H]
and (2) the actual data of the [Fe/H] distribution.  The resultant
index distributions are remarkably consistent with the observations.
As the process should be in principle reversible (Papers II, III, and IV), 
we also try {\it inverse-transformations} from index distributions to metallicities in Section 6.
This yields metallicity distributions of similar shapes to the observation. 
Section 7 discusses the implications of our findings.

\section{THE M31 GLOBULAR CLUSTER SAMPLE}

We use the dataset of confirmed M31 GCs by \citet{caldwell09,
caldwell11} who obtained high-S/N GC spectra with multi-fiber
spectrograph Hectospec \citep{fabricant05}, on the 6.5 m Multiple
Mirror Telescope (MMT) during 2004--2007. The spectra cover the
wavelength range of 3700--9200 {\small \AA} at 5 {\small \AA} resolution, and the
median S/N at 5200 {\small \AA} of the main set of spectra is 75 per {\small \AA},
with some as high as 300. Lick indices measured using the passbands
defined by W94 and Worthey \& Ottaviani (1997) are
calibrated on the Lick system (as redefined by Schiavon (2007, hereafter S07))
using Lick standards. For further details on the dataset, including
sample selection, observation and reduction, we refer the reader
to \citet{caldwell09, caldwell11} and \citet{schiavon12}. Of the
316 old GCs ($>$\,6 Gyr) reported by \citet{caldwell11}, we are primarily
interested in the clusters older than 10 Gyr with high-quality
measurements (S/N $>$\,20). Our final spectroscopic sample contains
280 GCs.

Recent observations reveal that M31 is a bulge-dominant galaxy
harboring an extended bulge \citep{hurley04, merrett06, lee08b}.
\citet{morrison04} suggested that there is a subsystem of old, metal-poor
GC with thin-disk kinematics in M31. 
However it was later pointed out that the majority of thin-disk GCs are 
in fact much younger than 10 Gyr and metal-rich \citep{beasley04}.
With a better classification of young and old clusters 
afforded by high quality spectra by \citet{caldwell09, caldwell11},
we consider our sample of old GCs as one belonging to the spheroid population of M31
and intend on including all of the GCs in our analysis.
The old, spheroidal component of M31 GCs used in this study is
analogous to GC populations in elliptical galaxies.

We supplement the spectroscopic analysis with color--magnitude
diagrams (CMDs) of individual stars in M31 GCs, based on 
{\it HST} photometry from literature. As was discussed in Paper I, contribution from HB stars
affect the strengths of absorption line indices, and the presence
of HBs in a cluster can be directly identified on its CMD.  
Those clusters whose HB morphology was classified as red or blue
based on CMD studies are employed in this work for a comparison
with spectroscopy. Eleven GCs with blue HB stars and 6 GCs with red HB
stars are taken from the combined list of GC CMDs by \citet{rich05}
and \citet{perina09, perina11}.

We use $ugz$-band integrated magnitudes by \citet{peacock10}
who performed photometry on archival SDSS images of M31 GCs. The
data provide 267 and 256 clusters in $g-z$ and $u-z$ colors,
respectively, of the 280 GCs in the spectroscopic sample. For
$V$-band magnitudes, we employ photometry published in the Revised
Bologna Catalog \cite[RBC v4.0,][]{galleti04}.  We also utilize the
far-ultraviolet (${\rm FUV}$, 1350--1750 {\small \AA}) and the near-ultraviolet
(${\rm NUV}$, 1750--2750 {\small \AA}) magnitudes of the clusters taken
from \citet{kang12} and \citet{rey07} who obtained integrated
photometry for 418 and 257 M31 GCs and GC candidates in ${\rm FUV}$
and ${\rm NUV}$, respectively, with {\it Galaxy Evolution Explorer}
({\it GALEX}).  Combined data of {\it GALEX} ${\rm UV}$ mags and
optical mags can offer a clue to HB temperature structure in stellar
populations.  The cluster sample is dereddened by adopting an average
$E(B-V)$ estimated from the lists of reddening values by \citet{barmby00},
\citet{fan08} and \citet{caldwell11} (see also \citet{kang12}). 
We limit our analysis to the clusters with low extinction
$E(B-V)\,<\,0.16$ for ${\rm FUV}$, ${\rm NUV}$, and $V$ magnitudes
to avoid large photometric uncertainties caused by M31 internal
reddening.  The number of lower extinction GCs having photometry
common in ${\rm FUV}$, ${\rm NUV}$, and $V$ magnitudes, as well as
spectroscopic information is 62.

\section{THE NONLINEAR NATURE OF COLOR--METALLICITY AND INDEX--METALLICITY RELATIONS}

Early investigation suggested that the presence
of hot blue HB stars manifests itself by making integrated
CMRs and index--metallicity relations (IMRs) {\it nonlinear} \citep{lee00,
lee02}. The use of ACS data enabled the nonlinear CMR
issue to be closely examined on observational grounds.
\citet{peng06} reported a notable nonlinearity in their empirical
CMR using $g-z$ colors and spectroscopic measurements of [Fe/H] for
GCs in the Milky Way, M49, and M87.  Paper I reproduced
the observed $g-z$ CMR by \citet{peng06} using a coeval group of
old (13 Gyr) model clusters.  Paper I, in essence, describes that the quasi-inflection point 
in CMRs has the effect of projecting the uniformly 
spaced metallicity points onto larger
color intervals, and thus can produce bimodal GC color distributions
from a broad [Fe/H] distribution, even if it is unimodal. 
The scenario gives simple yet cohesive explanation to the key observations, 
including the presence and properties of bimodality in color distributions 
and its intimate link to the host galaxy luminosity.

Figure 1 presents our theoretical CMRs and IMRs.
The stellar population simulations are
based on the Yonsei Evolutionary Populations Synthesis (YEPS) model
\cite[Papers I, II, III, and IV;][]{chung13}.  Predictions of absorption-line
indices are constructed based on the widely used polynomial fitting
functions parameterized in terms of {$T_{\rm eff}$}, log $g$, and [Fe/H].
YEPS models used in this study are computed with the fitting functions 
by Johansson \etal\ (2010, hereafter J10), that are based on the MILES stellar library \citep{sanchez06}.
We adopt 12 Gyr old model clusters with the [$\alpha$/Fe] = 0.14 \citep{puzia05, chung13}.
The solid lines represent model predictions
that account for systematic HB star contribution. 
The dashed lines, on the other hand, represent models without a
prescription for HB stars, for which post-RGB stars are absent.
A test shows that adding a fixed HB type regardless of GC metallicity 
affect only the zero-point of colors and index strengths as expected. 
Such variations do not affect our conclusion drawn from Figure 1, as well as Figures 2 and 3. 
The YEPS model predictions using J10 fitting functions are given in Tables 1--5 
for the indices Mg{\it b}, $\langle$Fe$\rangle$, H$\beta$, H$\gamma_{F}$, and H$\delta_{F}$
with and without the HB prescription.

The models in Figure 1 suggest that, because integrated light of main-sequence
and RGB stars behave nonlinearly as a function of metallicity at
given ages, the models without HB stars (dashed lines) exhibit some
measure of nonlinearity. 
The wavy feature in the CMRs and IMRs is greatly enhanced when the HB
effect is included (solid lines), increasing the sensitivity to
metallicity at [Fe/H] $\simeq$ $-1$.
For comparison, we overplot the observed $g-z$ and
$u-z$ colors \citep{peacock10} and the observed H$\delta_{F}$ and
Mg{\it b} indices as functions of [Fe/H] \citep{caldwell09, caldwell11}. 
The values of [Fe/H] are derived from a bi-linear relation between $\langle$Fe$\rangle$ 
and [Fe/H] for the Milky Way GCs (Caldwell \etal\ 2011, see their Figure 8).
The $g-z$ and $u-z$ colors respectively have 267 and 256 matched
GCs with known spectroscopic [Fe/H].
Unfortunately, colors of M31 GCs are much more prone to uncertainties caused by extinction than
spectral indices, showing too much scatter in the data to discern
any meaningful relationship in comparison with the models. 
In contrast, the relations for H$\delta_{F}$ and Mg{\it b} are better defined by the data
and they are reproduced reasonably well by the models.

It is clear in Figure 1 that
the shape and slope of CMRs and IMRs depend on the choice of colors and
indices due to their varying sensitivities to abundance and temperature. 
For the $g-z$ color (upper left panel), the models with HBs are bluer by 0.1 mag at most.  
Balmer index H$\delta_{F}$ (upper right), 
being one of the most temperature sensitive indices,
is a prominent tracer of HB stars in a stellar population, 
thus displaying a much stronger wavy feature in the relation. 
The increase in line strength reaches up to 1.5 {\small \AA} compared to the model
prediction for H$\delta_{F}$ without HB star inclusion. 
The relative amount of shift brought on by HB stars in $g-z$ and H$\delta_{F}$
corresponds to almost 4 times higher sensitivity to HBs for H$\delta_{F}$ 
considering the length of the baselines. 
Interestingly enough, HB stars clearly also affect well-known metallicity indicators, 
$u-z$  and Mg{\it b} (bottom panels).  
The characteristics of $u$-band colors
as metallicity tracer were explored in depth by \citet{yi04} and Paper II. 
The $u-z$ CMR displays less inflection in comparison with
that of $g-z$, because $u$-band colors are relatively insensitive
to temperature variation due to the Balmer Jump where $u$ band is located. 
Because the metal-line indices trace more directly the elemental abundance than colors do, 
the model predicts a relatively weaker wavy feature along the [Fe/H]--Mg{\it b} relation.
The relative HB sensitivity of Mg{\it b} is about a half of that of $u-z$
considering the length of the baselines.
However, this result demonstrates that the effect of HB stars on Mg{\it b} is certainly non-negligible.
Therefore greater caution is required in deriving cluster metallicity
directly from Mg{\it b} (see also Section 6 for more details).

\section{THE NONLINEAR INDEX--INDEX RELATIONS AND HORIZONTAL BRANCHES}

Figure 2 shows the distributions of 280 old, spheroidal component of M31 GCs 
\citep{caldwell09, caldwell11} in the planes of Balmer indices (H$\beta$ and H$\delta_{F}$) 
against the metal indices (Mg{\it b} and $\langle$Fe$\rangle$). 
The observed data define highly nonlinear index--index relations
and exhibit a division into two groups: 
(1) the weaker metal-line group with stronger Balmer-lines 
and (2) the stronger metal-line group with weaker Balmer-lines. 
The division occurs at Mg{\it b} $\simeq$ 2.0, $\langle$Fe$\rangle$ $\simeq$ 1.5, 
H$\beta$ $\simeq$ 1.8, and H$\delta_{F}$ $\simeq$ 1.4, respectively. 
We overlaid our models that do not include the HB prescription (Tables 1--5). 
The metal-rich group of GCs has weaker Balmer lines 
and is better explained by old (12, 13, and 14 Gyr) model lines than younger lines,
albeit a slight offset for H$\delta_{F}$.
On the other hand, the metal-poor GCs with stronger Balmer lines 
are predicted to have younger (7, 8, and 9 Gyr) ages. 
As a result, the models without HBs invoke the age difference 
between the metal-poor and rich groups by $\gtrsim$\,6 Gyr, 
with the metal-rich group being older. 
This seems inconsistent with the general notion of the age--metallicity relation 
that clusters with higher metallicities formed from more processed gas 
and thus are on average younger.

Figure 3 gives the single-age models that incorporate the HB prescriptions
(solid lines, Tables 1--5). 
The overlaid model predictions (red solid, dotted lines) in the upper four panels 
represent the models based on the J10 fitting functions\footnote{The model absorption indices generated with the J10, S07 and W94 fitting functions are available online at http://web.yonsei.ac.kr/cosmic/data/YEPS.htm.} .
Compared to cool stars, hotter stars have stronger Balmer-lines 
(reaching peaks at $\sim$10,000 K) and weaker metal-lines. 
The models predict that only metal-poor group contains hot ($>$\,8,000 K) HB stars, 
which leads to stronger Balmer lines and weaker metal lines 
than the models without HB stars (dashed lines). 
As a consequence, the model with HB stars features the wavy relations,
reproducing the observed index--index behaviors. 

In order to check whether the nonlinear feature depends on a particular choice of fitting functions, 
similar sets of simple stellar population (SSP) models are generated 
adopting two different sets of fitting functions presented by S07 and W94
and shown in the lower four panels of Figure 3.
The model adopts the S07 fitting functions for all indices, that employed
a more recent, superior spectral library \citep{jones99} than W94. 
While the overall agreement between
the two different sets of fitting functions are found to be good
for most indices \cite[S07;][]{chung13}, S07 fitting functions for
Balmer lines yield a better match to observations at low metallicity
than those of W94.  
For the limited cases of Mg{\it b} and $\langle$Fe$\rangle$, however, 
W94 is used, whose Mg{\it b} and $\langle$Fe$\rangle$ fitting functions
render more sensitivity to the variation of [Fe/H] in the high temperature range of HB stars (6000 -- 12000 K)
and thus produce more inflected IMRs for these indices. 
The stellar population model based on the fitting functions by S07 and W94 
also shows the wavy feature and reproduce the observed index--index relations very well.

Our theoretical rendition of M31 GC spectroscopy is supported by recent photometric studies. 
As our theoretical evidence asserts that HB stars are the main driver 
behind nonlinearity in index--index relations, 
the direct identification of HB stars on the CMDs of metal-poor GCs
is confirmation of the model prediction. 
Individual stars in M31 GCs resolved by {\it HST} reveal HB morphology 
into the blue HB and the red HB through inspection of the CMDs  \citep{rich05, perina09, perina11}. 
It is shown in Figure 3 that 11 metal-poor GCs with enhanced Balmer indices 
possess hot ($>$\,8,000 K) HB stars (blue squares), 
whereas six higher-metallicity GCs with weak Balmer lines do not (red squares).
Further support for the presence of hot HB stars in metal-poor group of GCs in M31
comes from the integrated UV light by the {\it GALEX} space telescope \citep{kang12}. 
Hot HB stars are dominant UV sources in old GCs \citep{oconnell99}, 
and 27 metal-poor GCs with strong Balmer indices 
indeed show strong UV light ($NUV-V\,<\,3.5$ and $FUV-V\,<\,4.5$), 
corroborating the presence of hot HB stars in the metal-poor GCs (cyan squares). 
Three UV-weak ($NUV-V\,>\,4.5$ and $FUV-V\,>\,5.0$) GCs are identified (orange squares), 
one of which is a confirmed red-HB GC from its CMD (red squares). 
They all are located in the metal-rich region.

\section{THE METALLICITY--INDEX NONLINEARITY AS ORIGIN OF INDEX BIMODALITY}

As shown in Figures 2 and 3, the sample of 280 old, spheroidal component GCs of M31 
exhibit a division into two groups. 
This Section examines how the nonlinear IMRs 
bring about the spectroscopic dichotomy of M31 GCs. 
Any viable scenario for the origin of the GC dichotomy should explain 
(1) the bimodality both in the metal-index and the Balmer-index distributions, 
and (2) the cause of the sharp difference in bimodality strength
between metal- and Balmer-lines. 

Figures 4$a$ and 4$b$ illustrate the process of conversion from the intrinsic metallicities to the indices 
via our theoretical metallicity--index relations.  
As in Figure 3,  Figures 4$a$ and 4$b$ represent
the models generated with the different fitting functions. 
Red model lines in Figure 4$a$ are based on J10 fitting functions 
and the orange model lines in Figure 4$b$ are of S07 and W94 fitting functions.
For the underlying [Fe/H] spread, 
we make a simple assumption of a single Gaussian distribution 
of $10^6$ model GCs (top rows of Figures 4$a$ and 4$b$) to avoid small number statistics.
The assumed mean and standard deviation of the Gaussian [Fe/H] function 
are respectively $-0.9$ and 0.6, and $-1.0$ and 0.5, 
which best reproduce the observed index distributions in the bottom rows of Figures 4$a$ and 4$b$. 
Observational uncertainties are taken into account in the simulation of the index distributions.
The inflection in our theoretical [Fe/H]--index relations is visible 
not only for Balmer indices but also for Mg{\it b}.
Such nonlinear feature has the effect of projecting the equi-distant metallicity intervals 
onto broader index intervals, 
causing scarcity in the index domain near the quasi-inflection point on each [Fe/H]--index relation. 

As an aid to visualizing the simulated index distributions, 
we plot the indices of model GCs against their mass as an independent parameter (second rows of Figures 4$a$ and 4$b$). 
The divide between two vertical bands of GCs is visible 
and agrees well with the observed data. 
The resulting index histograms show bimodality both for Mg{\it b} and for Balmer indices 
with the scarcity at each quasi-inflection point reflected as dips (third rows of Figures 4$a$ and 4$b$).
Given that the number of observed GCs is much smaller than that of the modeled ones, 
we repeat random sampling of 280 GCs from the $10^6$ model GC sample
in order to assess the reliability of frequency of GCs in each bin. 
The gray shaded region in the index histograms (third rows of Figures 4$a$ and 4$b$)
represent the estimated confidence intervals---one standard deviation of the 10,000 repeated samplings.
Therefore we find good agreement between the theoretical predictions and the observations (bottom rows of Figures 4$a$ and 4$b$).

We perform mixture modeling analysis for a quantitative comparison between the distributions of 280 observed GCs and 10,000 randomly selected model GCs using GMM code (Gaussian Mixture Modeling, Muratove \& Gnedin 2010).
In addition to the general method of calculating the likelihood of a given data belonging to a mixture of two Gaussians (expressed in terms of $P(\chi^2)$, Ashman et al. 1994). 
GMM provides further statistical tests of bimodality in forms of $DD$, a measure of separation between peaks relative to their widths, and kurtosis, a measure of peakedness of a given distribution.
The kurtosis statistic assesses whether a distribution is bimodal such that a negative kurtosis corresponds to a more flattened shape of the sum of two populations. 
Note that {\it kurt}$\,<\,0$ is a necessary but not sufficient condition of bimodality. 
Table 6 presents the resulting GMM outputs.
Bimodality is preferred for Mg{\it b} distributions for observed and simulated GCs ($P(\chi^2) = 0.001-0.01$).
The fraction of GCs assigned to the metal-rich group is similar for the observation and simulations (i.e., $59\,\% \sim 64\,\%$). 
The observed and the simulated distributions of H$\beta$, H$\gamma_{F}$, and H$\delta_{F}$ are shown 
to be bimodal at the $99.9\,\%$ level, displaying two distinct peaks.
For all three Balmer lines, the metal-poor peaks are slightly more dominant 
with the simulations consistently predicting lower fractions of metal-rich GCs compared to the observation.
On the whole, we find good agreement between the observation and the models.

We note that the offset between the models and the observation are translated to the shift in peak locations in the simulated index distributions.
Our stellar population models show that, for given input parameters, 
the absolute quantities of output are rather subject to the choice of model ingredients 
such as stellar evolutionary tracks, flux libraries, and fitting functions, 
and the different choices can result in up to $\sim$\,0.4 {\small \AA} variation 
in Mg{\it b} and Balmer strengths.
Hence, we put more weight on the {\it relative} index values, 
i.e., the number ratios between index-weak and strong GCs 
and the overall morphologies of the simulated index histograms.
The recent work by \citet{chung13} discusses 
the systematic effects exerted by varying MDFs and ages, 
as well as the use of different stellar population models 
on the projected index distributions (see their Figures 20--23). 
The cases for different choices of model ingredients are also addressed 
in Section 3 of the same paper (see their Figure 5).

In Figure 5, we apply the identical projection scheme 
to the {\it actual} data of the [Fe/H] distribution comprised of 280 old GCs in M31. 
Using the observed [Fe/H] distribution, we examine the index distributions produced 
by conventional linear fit and by projection through the model relations.
The observed [Fe/H] distribution (the grey histograms along the $y$-axes of all panels) 
is best described by a single Gaussian with the mean and standard deviation of $-1.05$ and 0.57, 
which agree well with those of the modeled [Fe/H]'s in Figure 4. 
It is noteworthy that the observed, unimodal [Fe/H] histogram is at odds with the popular view, 
in which GC systems consist of two groups with different metallicities. 
With such an underlying metallicity spread, 
the indices converted via the conventional linear fit to the observed [Fe/H] versus index data 
are bound to have unimodal index distributions (second row), 
and thus inconsistent with the observations. 
In contrast, our theoretical [Fe/H]--index relations produce clear bimodality 
in the Balmer and Mg{\it b} index distributions for both models with J10 fitting functions (third row) 
and the models with S07 and W94 (bottom row). 

Table 7 lists our GMM analysis of Figure 5.
The observations and the non-linear transformations through two model sets 
are bimodal with GMM probabilities of $99.9\,\%$. 
For Mg{\it b}, sharper metal-poor peaks are reflected in somewhat less significant $P({\it kurt})$ values.
Unlike in Figure 4, the simulations predict higher fractions of metal-rich GCs compared to the observations. 
In the case of linear transformations, GMM finds the probabilities of rejecting unimodal Gaussian 
to be insignificant ($P(\chi^2) = 0.46-0.47$).
Most of the GCs ($>\,94\,\%$) are assigned to either the metal-poor group (Mg{\it b}) or the metal-rich group (Balmer lines).

\section{THE INDEX--METALLICITY NONLINEARITY AND SPECTROSCOPIC METALLICITIES}

Previous spectroscopic studies of extragalactic GC systems have made use of 
metal-line indices such as Mg{\it b} and a composite [MgFe]$^\prime$ index 
as proxies to cluster metallicity \cite[e.g.,][]{strader07,woodley10}.  
In case of NGC 5128 GC system, for instance, \citet{woodley10} found
that bimodality is preferred for their measured [MgFe]$^\prime$.  
Because [MgFe]$^\prime$ index is obtained directly from high-quality spectra 
of NGC 5128 GCs and known to be almost independent of [$\alpha$/Fe], 
they concluded that their bimodal [MgFe]$^\prime$ distribution 
indicates the existence of two distinct metallicity subpopulations.  
However, we have shown that the strengths of absorption line indices  
are affected in varying degrees by hot HB stars in old stellar populations 
according to their respective responses to HB temperature.
We have subsequently demonstrated both theoretically and observationally
that the nonlinearity in the [Fe/H]--index relations can generate
bimodal index distributions, even for the Mg{\it b} index, 
from unimodal metallicity spread.  

More importantly, Balmer lines exhibit very strong bimodal index distributions, 
which are translated routinely into bimodal metallicity distributions. 
Most studies have derived spectroscopic [Fe/H]'s 
based {\it jointly} on metal-lines and Balmer lines: finding best-fitting age, metallicity, and [$\alpha$/Fe]  
through multivariate fits to SSP models and/or an iterative method using 
the grids of Balmer lines and metal-sensitive indices 
 \cite[e.g.,][]{puzia05, cenarro07, beasley08, park12}. 
Therefore, Balmer lines have played an important role in establishing the notion of bimodality in spectroscopic metallicity.
In light of this, the previous view of the absorption indices being {\it linear} metallicity tracers requires 
modification \citep{chung13}, and a revisit to spectroscopic metallicity measurements of GCs may prove worthwhile.

Figure 6 attempts at obtaining metallicity distributions by inversely transforming each individual index distributions to metallicities.
The process of inverse conversion involves complication arising from the incompleteness of the current models, 
in that the visible offsets between observations and models result in the slight offsets in the peak positions.
Considering the caveat, the model incompleteness allows room for displacement of the models within a few tenth of an {\small \AA} ($<\,\left|0.4\right|$ {\small \AA}), for more accurate simulations of MDFs.
We displace the model lines in both directions along x-axis ($<\,\left|0.4\right|$ {\small \AA}) in steps of 0.01 {\small \AA}, then determine the best-fit $\chi^2_{min}$ to the observed data. Gaussian kernel smoothing is applied to the observation (top row) in order to reduce the stochastic variations. 
These kernelled data of observed index distributions are transformed via the models that are now displaced accordingly to the amount determined for each index.
The red solid lines and the orange dahsed lines in the second row correspond to the models based on fitting functions of J10 and fitting functions of S07 and W94, respectively, 
and the resulting MDFs are shown as red and orange histograms in the fourth and bottom row.
The MDFs derived from Mg{\it b} exhibit broad unimodal distributions.
The MDFs inferred from Balmer lines show similar unimodal shapes 
with somewhat extended metal-rich wings, which appears consistent with the observed MDF in Figure 5.
We note that, in contradiction to the visual impression, the values of $P(\chi^2)$ indicate a 
high probability of bimodality for the inferred MDFs (Table 8).
\citet{muratov10} discusses the limitations of the $P(\chi^2)$, and states that ``GMM is more a test of Gaussianity than of unimodality".
The $P(DD)$ and $P({\it kurt})$, however, validate the preferred unimodal distributions over bimodal ones.
On the contrary to the above cases, linear conversions (blue dotted lines in the second row) preserve the morphology of each index distribution, and exhibit strong bimodality in inferred metallicity distributions  (blue histograms in the third row), which is inconsistent with the observation.
This issue entails far-reaching implications
and will be fully explored in a forthcoming paper (S. Kim et al. 2013, Paper VI in preparation).

\section{DISCUSSION}

We show that the spectroscopic dichotomy of the M31 GCs emerges 
not necessarily because of any division in physical parameters of GCs 
such as metallicity and age. 
Instead, the phenomenon is most likely attributed to the difference 
in the hot HB proportion of stellar populations, 
which is a strong nonlinear function of GC metallicity 
and increases abruptly in metal-poor GCs.
The large hot star fraction of metal-poor GCs 
simultaneously makes metal-lines weaker and Balmer-lines stronger 
than they otherwise might be.  
As a result, GCs that have the index values around the midpoint of distributions 
are relatively scarce, which explains the bimodality in index distributions.
The results that are based on spectroscopy---a more detailed probe into
stellar contents in GCs---are consistent with the view 
that the nonlinear metallicity-to-color transformation 
is responsible for photometric color bimodality of GCs (Papers I, II, III, and IV).


Perhaps the strongest argument against our scenario 
for the spectroscopic division of M31 GCs comes from our Milky Way, 
which is a similar disk galaxy and 
serves as a prime example of confirmed two GC groups present \citep{harris06}. 
Despite many similarities shared by the two galaxies, however, 
there is increasing evidence that the spheroidal components of M31 and our Galaxy 
have distinct characteristics.  
For instance, extensive investigations into the kinematics of GCs and planetary nebulae 
suggest that M31 is a bulge-dominant disk galaxy, 
unlike the Milky Way \citep{hurley04,merrett06, lee08b}.  
These studies revealed a general trend of increasing velocity dispersion 
for M31 GCs with distance, indicating the presence 
of a system of pressure-supported, dynamically hot halo.  
Furthermore, spectroscopic analyses showed that 
metallicity distribution of old M31 GCs differs from that of the Galactic GCs 
in that the M31 GC system lacks bimodality in its metallicity distribution \citep{caldwell11,galleti09}. 
Therefore, M31's spheroid does not seem to have the same GC formation history 
as that of the Milky Way 
and, the two galaxies do not necessarily share the common origin of the GC division. 

Hierarchical models for galaxy formation depict the shaping of 
a single massive galaxy through merging of thousands of lower-mass building blocks. 
This picture may leave little room for the possibility of a GC system 
containing {\it merely} two subpopulations.
Indeed, the continuous nature of the physical parameters of old GCs in M31
can stem from its virtually continuous chemical evolution early on ($\sim$\,12 Gyr ago). 
The chemical enrichment seems to have been achieved on a relatively short timescale 
via many successive rounds of early, vigorous star formation in M31 halo's history.  
Interestingly, the old, spheroidal component of the M31 GC system used in this study 
is associated chiefly with the galaxy's bulge, 
and hence can be viewed as an analogy to GC systems 
belonging to typical elliptical galaxies.  
Large elliptical galaxies were the sites 
where the GC color bimodality was first discovered, 
and ever since ellipticals have been believed to harbor two groups of GCs with different genesis.
Our findings challenge such conventional wisdom, 
and the new insight into the structure of GC systems 
greatly simplifies theories of galaxy formation.

\acknowledgments
S.-J.Y. acknowledges support by Mid-career Research Program (No. 2012R1A2A2A01043870) 
through the National Research Foundation (NRF) of Korea grant funded 
by the Ministry of Education, Science and Technology (MEST), 
by the NRF of Korea to the Center for Galaxy Evolution Research (No. 2012-8-1743), 
and by the Korea Astronomy and Space Science Institute (KASI) Research Fund 2011 and 2012. 
This work was partially supported by the KASI--Yonsei Joint Research
Program (2012--2013) for the Frontiers of Astronomy and Space Science.
S.-J.Y. and S.K. would like to thank Daniel Fabricant, Charles Alcock, Jay Strader, Nelson Caldwell, Dong-Woo Kim, Jae-Sub Hong 
for their hospitality during their stay at Harvard-Smithsonian Center for Astrophysics in 2011--2012. 
S.-C.R. acknowledges support from Basic Science Research Program (No. 2012-003097) 
through the NRF of Korea funded by the MEST.


\clearpage
\begin{figure*}
\begin{center}
\includegraphics[width=18.5cm]{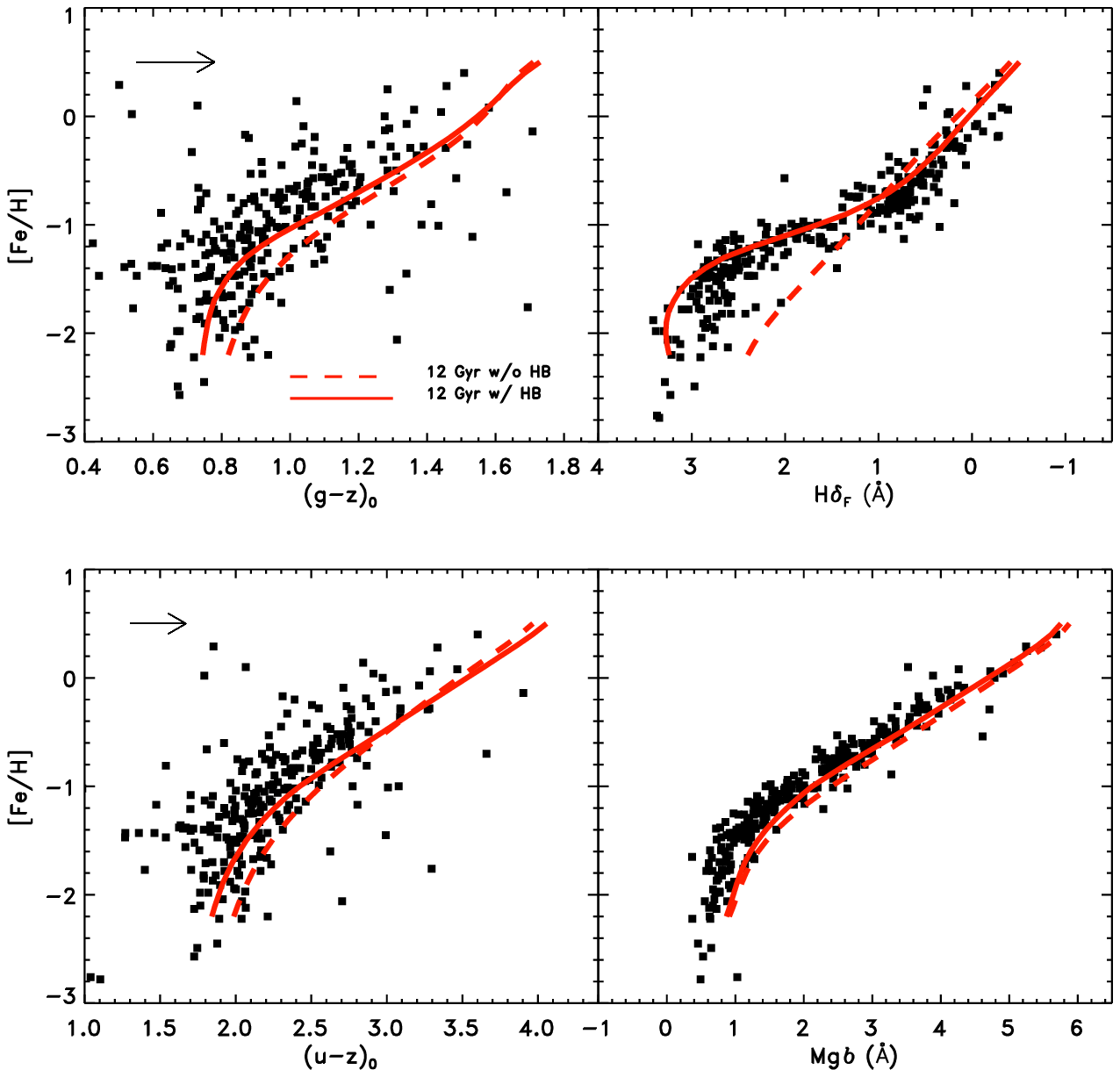}
\end{center}
\end{figure*}

\clearpage
\begin{figure*}
\begin{center}
\caption{
Empirical and theoretical color-metallicity relations and
index-metallicity relations.  Red curves are the 12 Gyr old Yonsei
Evolutionary Populations Synthesis (YEPS) model (Paper II; Chung
\etal\ 2012) depicting models including the prescriptions
for HB stars. Blue lines represent the 12 Gyr old model clusters
without inclusion of HB stars.  The theoretical CMRs are a fifth-order
polynomial fit to the model data.  The $\alpha$-element enhancement
parameter, [$\alpha$/Fe], is assumed to be 0.14 (Puzia et al. 2005).
Photometry for $u$ (ABMAG), $g$ (ABMAG), and $z$ (ABMAG) mags for
GC candidates in the left panels are from SDSS catalog (Peacock et
al. 2010). Using the archival images, Peacock et al. (2010) performed
photometry for the candidate clusters listed in Revised Bologna
Catalog (RBC v4.0, Galleti et al. 2004).  The averaged $E(B-V)$
from \citet{barmby00}, \citet{fan08} and \citet{caldwell11} are
adopted for dereddening. The arrow represents a reddening of $E(B-V)
= 0.1$.  267 and 256 matches against spectroscopically derived
[Fe/H] \citep{caldwell09, caldwell11} were found for $g-z$ and $u-z$
colors, respectively. For H$\delta_{F}$ and Mg{\it b}, 280 high
S/N, old clusters of M31 \citep{caldwell09, caldwell11} are used.
For a better comparison with color,  x-axis for H$\delta_{F}$ is
in reverse order.
\label{fig1}}
\end{center}
\end{figure*}

\clearpage
\begin{figure*}
\begin{center}
\includegraphics[width=17.3cm]{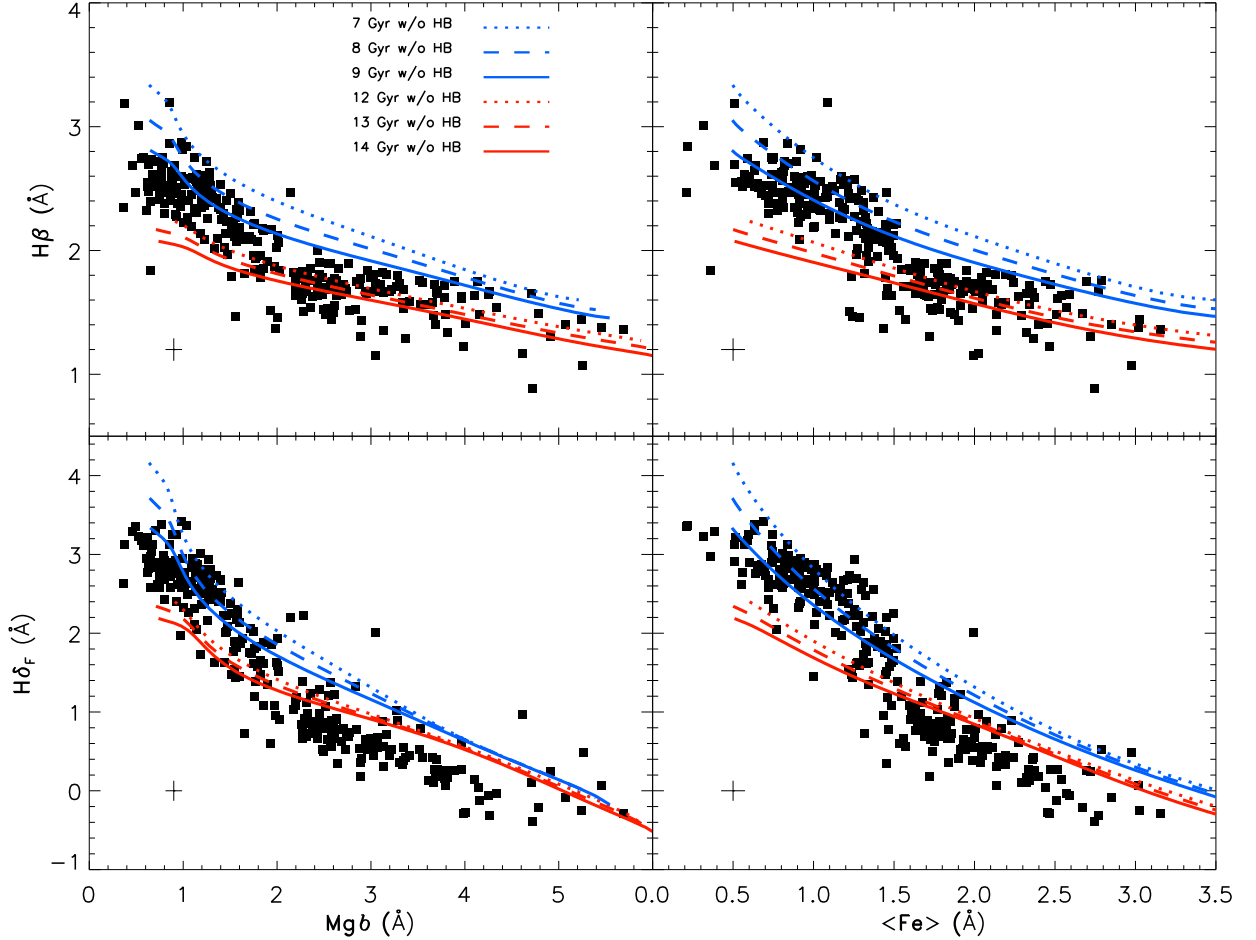}
\caption{
The distributions of 280 high S/N GCs (Caldwell et al. 2009, 2011) in the planes of Balmer indices (H$\beta$ and H$\delta_{F}$) against the metal indices (Mg{\it b} and $\langle$Fe$\rangle$) (a) the weaker metal-line group with stronger Balmer-lines and (b) the stronger metal-line group with weaker Balmer-lines. The division occurs at Mg{\it b} $\simeq$ 2.0, $\langle$Fe$\rangle$ $\simeq$ 1.5, H$\beta$$\simeq$1.8, and H$\delta_{F}$$\simeq$1.4, respectively. Overlaid YEPS models with varying ages that do not include the HB prescription are shown in blue for younger GCs (7, 8 , 9 Gyr) and in red for older GCs (12, 13, 14 Gyr). The models without HBs seems to invoke the age difference between the metal-poor and rich groups by as large as 6 Gyr.
\label{fig2}}
\end{center}
\end{figure*}

\clearpage
\begin{figure*}
\begin{center}
\includegraphics[width=15cm]{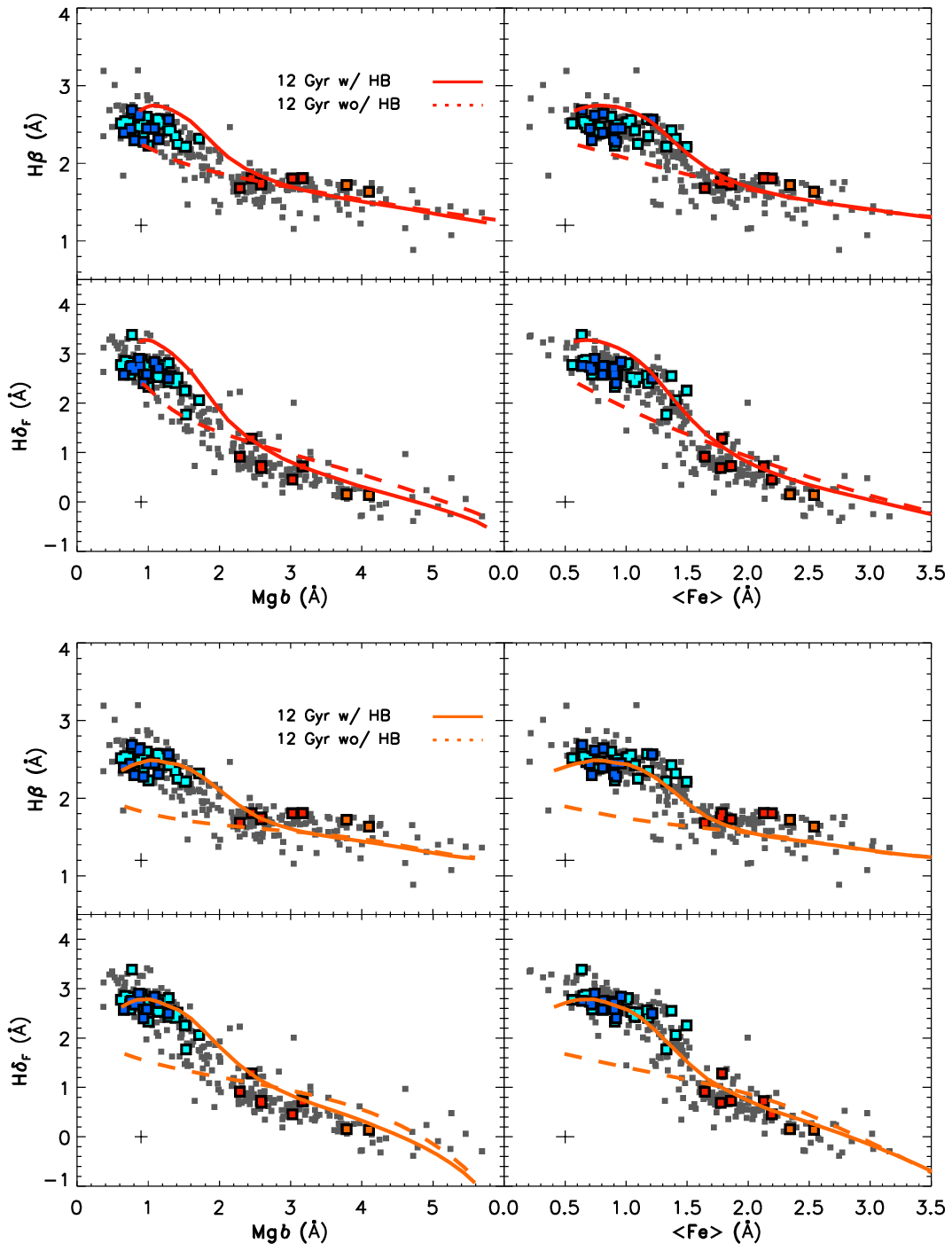}
\end{center}
\end{figure*}
\clearpage
\begin{figure*}
\begin{center}
\caption{
Same as Figure 2 but  the correlations between Balmer-line indices
(H$\beta$ and H$\delta_{F}$) and metal-line indices (Mg{\it b} and
$\langle$Fe$\rangle$) for M31 GCs are compared with the single-age
models with HB star inclusion (solid lines) and without HB star
inclusion (dashed lines). 
In the lower panel, the YEPS models use the functions presented by S07 for Balmer indices, while the models adopt W94 for metal line indices Mg{\it b} and $\langle$Fe$\rangle$ (orange solid and dotted lines).
Top left of each 2-by-2 panel: the H$\beta$--Mg{\it b}
relation for 280 GCs (gray squares). The error bars show the mean
observational uncertainties. 11 blue squares are GCs containing hot
($>$\,8,000 K) HB stars confirmed by their color-magnitude diagrams
with Hubble Space Telescope, and six red squares are GCs without
hot HB stars \citep{rich05, perina09, perina11}.
Cyan and orange squares represent 27 UV-strong ($NUV-V\,<\,3.5$ and $FUV-V\,<\,4.5$) and three
UV-weak ($NUV-V\,>\,4.5$ and $FUV-V\,>\,5.0$) GCs identified in the
{\it GALEX} UV photometric studies, respectively (Kang \etal\ 2012; Rey \etal\ 2007; RBC v4.0,
Galleti et al. 2004).  Five GCs are in common in the {\it HST} and
{\it GALEX} observations. The overlaid lines represent our model
predictions for 12 Gyr old GCs with the [$\alpha$/Fe] = 0.14 (Puzia
et al. 2005). 
Bottom left of each panel: same as ({\it Top left}), but
for the H$\delta_{F}$--Mg{\it b} relation.  Top right of each panel: same
as ({\it Top left}), but for the H$\beta$--$\langle$Fe$\rangle$
relation.  Bottom right of each panel: same as ({\it Top left}), but for
the H$\delta_{F}$--$\langle$Fe$\rangle$ relation.
\label{fig3}}
\end{center}
\end{figure*}

\clearpage
\begin{figure*}
\begin{center}
\includegraphics[angle=90, width=15cm]{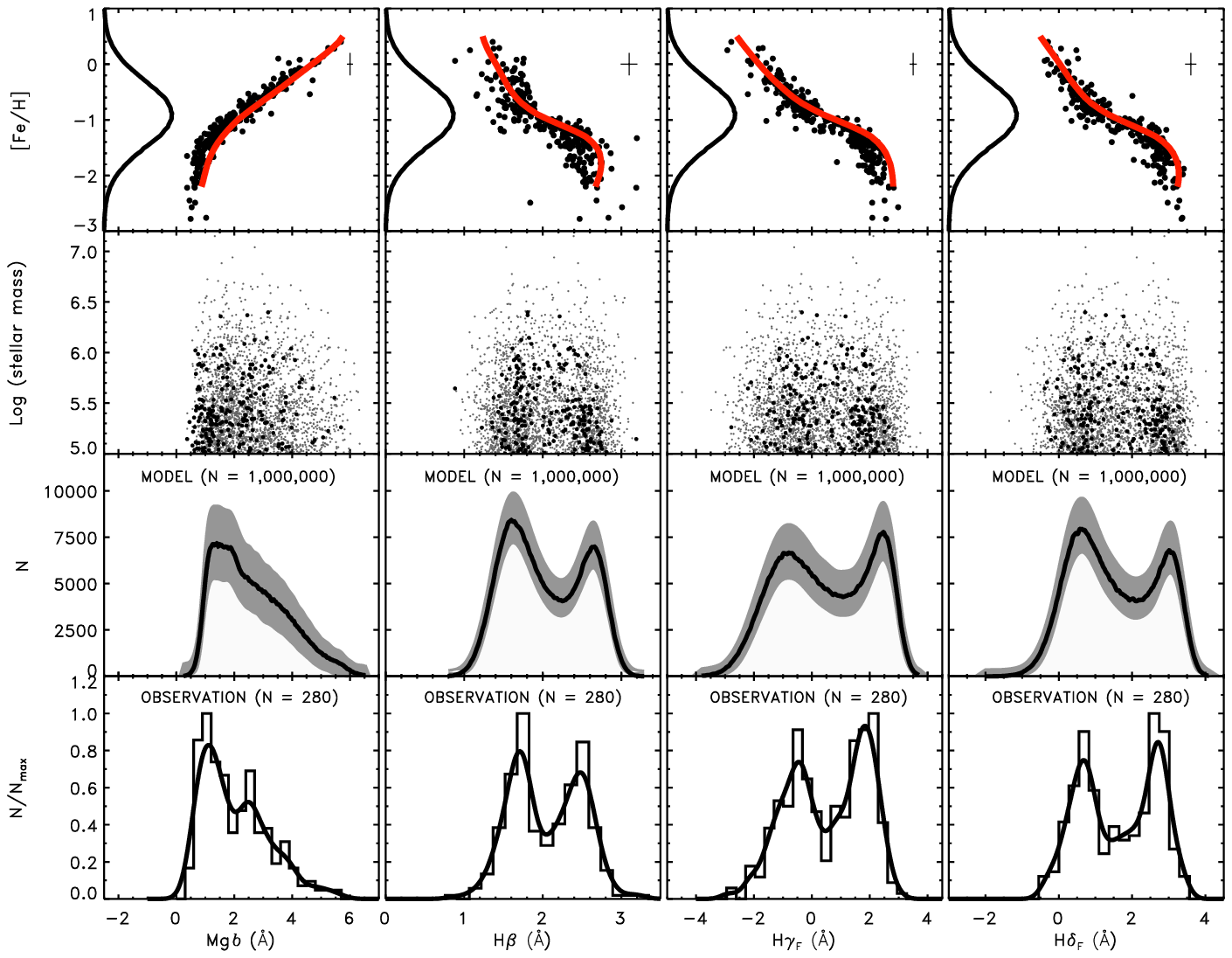}
\end{center}
\end{figure*}

\clearpage
\begin{figure*}
\begin{center}
\includegraphics[angle=90, width=15cm]{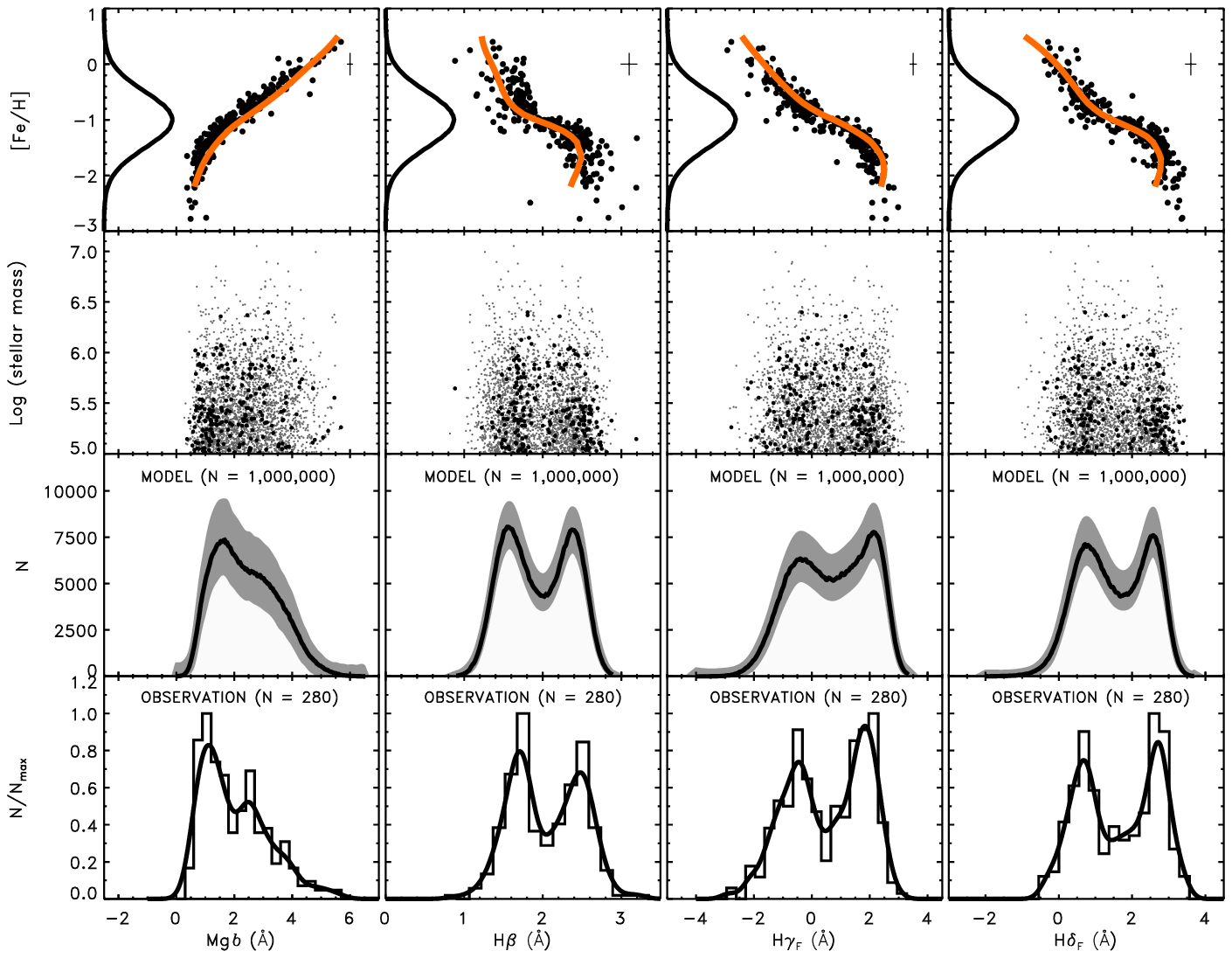}
\end{center}
\end{figure*}

\begin{figure*}
\begin{center}
\caption{$a$. Monte-Carlo simulations of the indices (Mg{\it b}, H$\beta$, H$\gamma_{F}$, and H$\delta_{F}$) distributions for $10^6$ model GCs.
Top row: the correlation between [Fe/H] and Mg{\it b}, H$\beta$ and H$\delta_{F}$ for the identical M31 GC sample in Figure 2, overlaid with the model with the HB prescription (red solid lines). A single Gaussian distribution along the y-axis represents the underlying metallicity distribution. Second row: indices vs. stellar mass diagram for 5000 randomly selected model GCs (gray dots) and the observed M31 GCs (black dots) (RBC v4). Indices are derived based on the metallicity distribution and theoretical index--[Fe/H] relations shown in top row. Stellar masses of GCs are derived from the integrated V-band absolute magnitudes (Caldwell et al. 2011), assuming a Gaussian luminosity function ($\langle$$M_{V}$$\rangle$ = $-6.72$ and $\sigma({M_V})$ = 1.5) for $M_{V}$. Third row: the resultant index distribution for $10^6$ model GCs. 
The bin sizes for each index distribution range from 0.01 to 0.03 {\small \AA}.
Observational uncertainties are taken into account in the simulation of the index distributions.
Given that the number of observed GCs is much smaller, 
280 GCs from the $10^6$ model GCs are drawn randomly to estimate the possible ranges of simulated index distributions.
The sampling is carried out 10,000 times for each index, 
and the grey shaded bands show the $1 \sigma$ range of individual bins.
The mean observational error is adopted for the bin size.
Bottom row: the observed index distributions for 280 GCs. The mean error is adopted for the bin width, and a Gaussian density kernel estimate is shown by solid line.
$b$. Same as Figure 4$a$ but using the models computed with S07 and W94 fitting functions (orange solid lines). 
\label{fig4}}
\end{center}
\end{figure*}

\clearpage
\begin{figure*}
\begin{center}
\includegraphics[width=17.3cm]{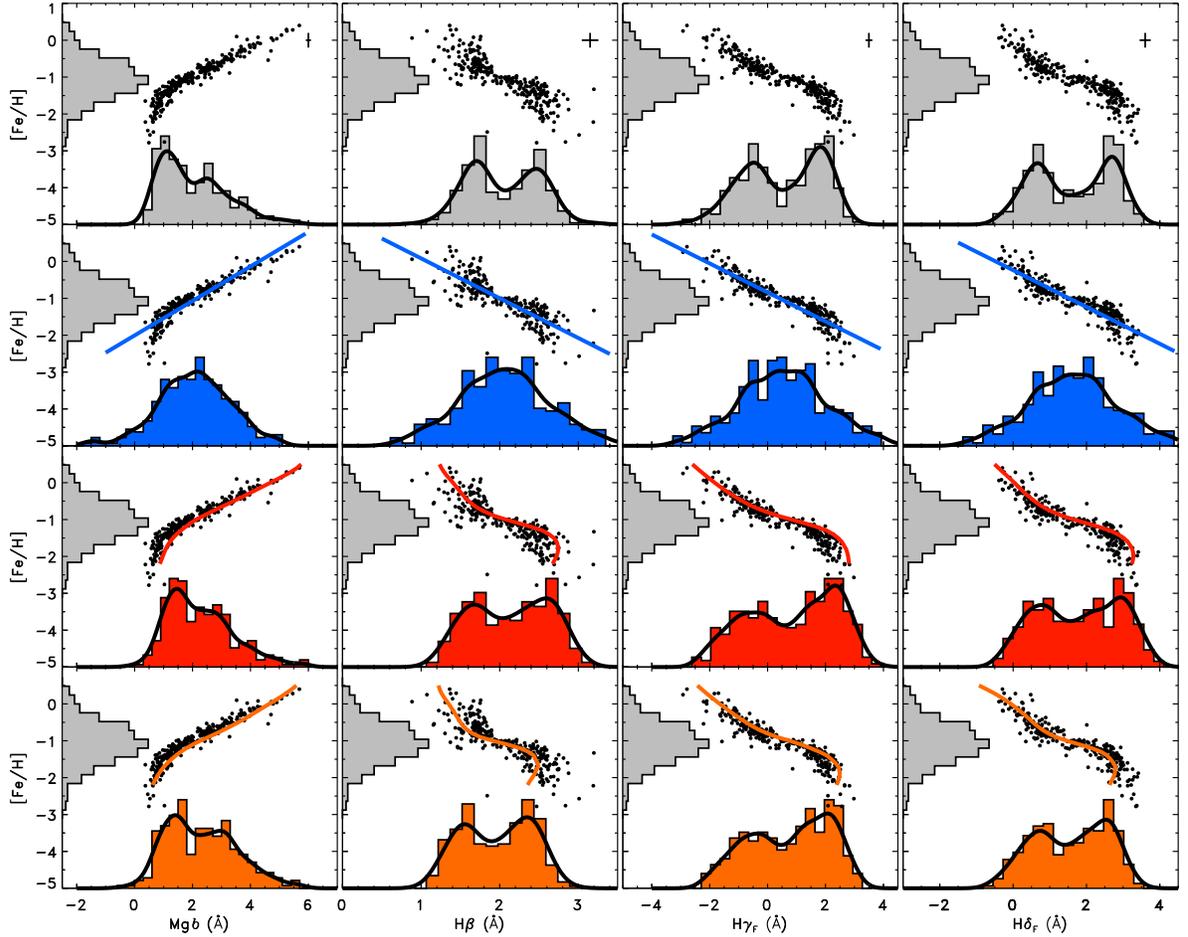}
\caption{
Comparison of simulated indices (Mg{\it b}, H$\beta$, H$\gamma_{F}$, and H$\delta_{F}$) distributions with the M31 GC observations. Top row: the index--[Fe/H] relations for the identical sample in Figure 2. The observed distributions of [Fe/H] and indices are shown along the y-axis and x-axis, respectively. Second row: the observed [Fe/H] histograms along the y-axes are the same as in the top row but the transformed index distribution (blue histogram) using a simple linear fit (blue solid lines) to the data. Third row: the transformed index distribution (red histogram) using our theoretical [Fe/H]--index relation (red solid lines). Bottom row: the transformed index distribution (orange histogram) using our theoretical [Fe/H]--index relation (orange solid lines) based on S07 and W94 fitting functions.
\label{fig5}}
\end{center}
\end{figure*}

\clearpage
\begin{figure*}
\begin{center}
\includegraphics[width=17.3cm]{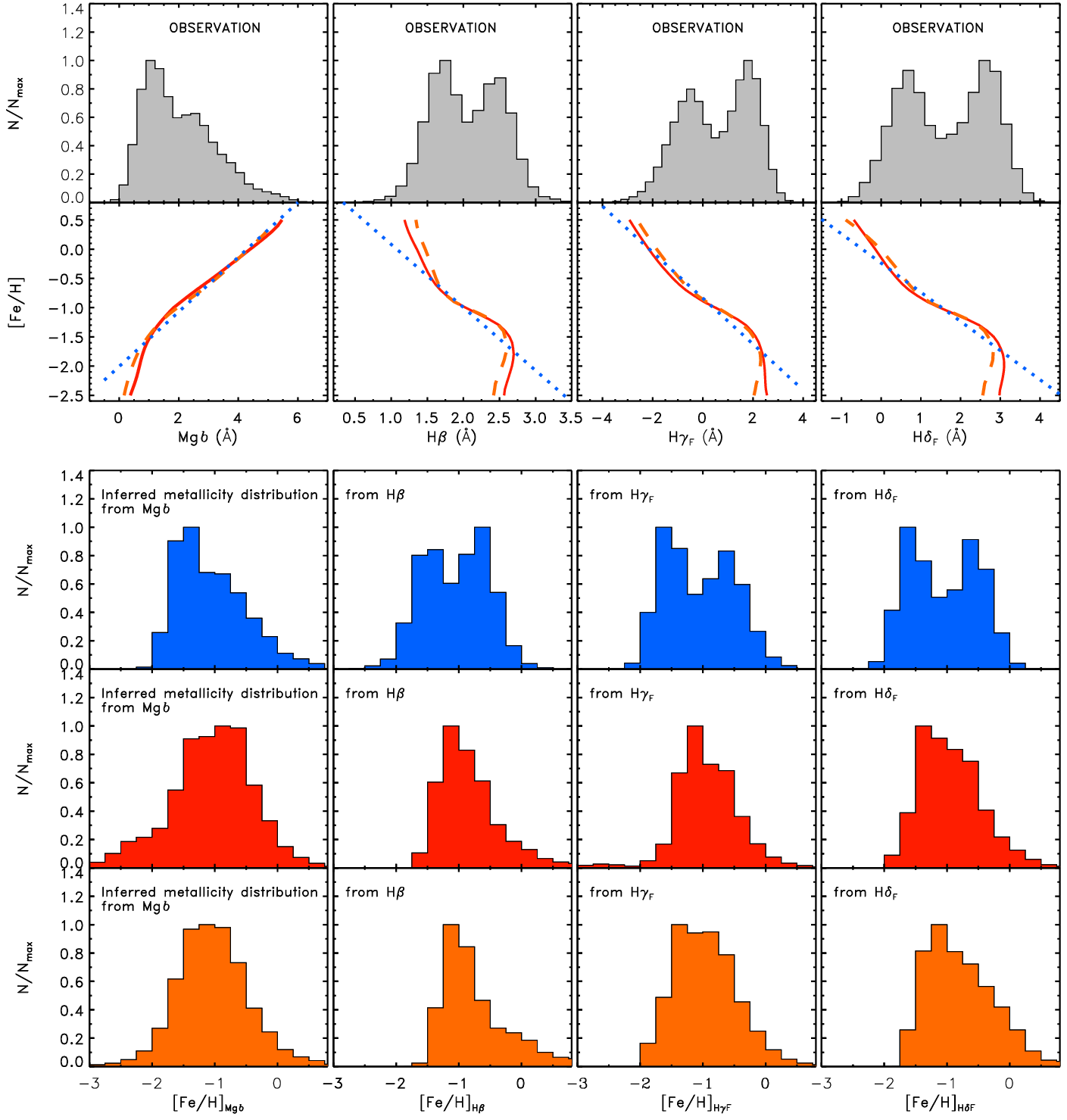}
\end{center}
\end{figure*}
\clearpage
\begin{figure*}
\begin{center}
\caption{
Mg{\it b}, H$\beta$, H$\gamma_{F}$, and H$\delta_{F}$ distributions and their inferred metallicity distributions for the M31 GC system.
Top row: Gaussian kernel smoothing is applied to the observed index distributions. 
Second row: 
color scheme is the same as in Figure 5.
Solid red lines represent theoretical [Fe/H]--index relations computed with J10 fitting functions displaced by --0.29 {\small \AA}, --0.05 {\small \AA}, --0.33 {\small \AA}, and --0.18 {\small \AA} for Mg{\it b}, H$\beta$, H$\gamma_{F}$, and H$\delta_{F}$, respectively.
Dashed orange lines represent theoretical [Fe/H]--index relations with S07 and W94 fitting functions offset by --0.31 {\small \AA}, 0.12 {\small \AA}, --0.19 {\small \AA}, and 0.04 {\small \AA} for Mg{\it b}, H$\beta$, H$\gamma_{F}$, and H$\delta_{F}$, respectively.
The blue dotted lines represent the linear fits to the kernelled data in the top row.
Third row: metallicity distributions obtained from each index, Mg{\it b}, H$\beta$, H$\gamma_{F}$, and H$\delta_{F}$ via linear transformations are shown as blue histograms. 
Fourth row: metallicity distributions obtained from each index using the models with J10 fitting functions are shown as red histograms. 
Bottom row: metallicity distributions obtained from each index using the models based on S07 and W94 fitting functions are shown as orange histograms.
\label{fig6}}
\end{center}
\end{figure*}

\clearpage
\begin{table}
\begin{center}
\scriptsize
\caption{The M$g{\it b}$ of the YEPS model GCs without ($w/o$) and with ($w$) HBs for [$\alpha$/Fe] = 0.14 and ages ($t$) of 7, 8, 9, 12, 13, and 14 Gyr. The models are computed using J10 fitting functions. \label{tbl-1}}
\begin{tabular}{ccccccccccccc}
\hline
[Fe/H] &
\multicolumn{12}{c}{Mg{\it b}} \\
\hline
{} &
\multicolumn{2}{c}{$t$ = 7} &
\multicolumn{2}{c}{8} &
\multicolumn{2}{c}{9} &
\multicolumn{2}{c}{12} &
\multicolumn{2}{c}{13} &
\multicolumn{2}{c}{14} \\
\hline
& $w/o$ & $w$ & $w/o$ & $w$ & $w/o$ & $w$ & $w/o$ & $w$ & $w/o$ & $w$ & $w/o$ & $w$ \\
\hline
--2.5 & 0.637 & 0.622 & 0.642 & 0.628 & 0.649 & 0.631 & 0.688 & 0.664 & 0.710 & 0.690 & 0.738 & 0.721 \\
--2.4 & 0.692 & 0.673 & 0.702 & 0.682 & 0.713 & 0.689 & 0.765 & 0.736 & 0.791 & 0.767 & 0.822 & 0.803 \\
--2.3 & 0.750 & 0.729 & 0.764 & 0.740 & 0.779 & 0.748 & 0.842 & 0.807 & 0.872 & 0.844 & 0.907 & 0.883 \\
--2.2 & 0.802 & 0.778 & 0.819 & 0.791 & 0.837 & 0.799 & 0.912 & 0.870 & 0.945 & 0.912 & 0.984 & 0.957 \\
--2.1 & 0.844 & 0.819 & 0.864 & 0.833 & 0.886 & 0.844 & 0.973 & 0.923 & 1.011 & 0.972 & 1.052 & 1.022 \\
--2.0 & 0.879 & 0.853 & 0.904 & 0.871 & 0.930 & 0.884 & 1.030 & 0.971 & 1.071 & 1.026 & 1.116 & 1.080 \\
--1.9 & 0.912 & 0.885 & 0.942 & 0.909 & 0.974 & 0.928 & 1.088 & 1.020 & 1.133 & 1.081 & 1.181 & 1.140 \\
--1.8 & 0.950 & 0.921 & 0.987 & 0.951 & 1.025 & 0.976 & 1.154 & 1.073 & 1.202 & 1.143 & 1.252 & 1.207 \\
--1.7 & 1.000 & 0.969 & 1.044 & 1.006 & 1.090 & 1.038 & 1.235 & 1.140 & 1.285 & 1.217 & 1.337 & 1.287 \\
--1.6 & 1.066 & 1.032 & 1.119 & 1.076 & 1.173 & 1.116 & 1.334 & 1.223 & 1.387 & 1.307 & 1.439 & 1.385 \\
--1.5 & 1.153 & 1.115 & 1.215 & 1.168 & 1.278 & 1.214 & 1.455 & 1.326 & 1.511 & 1.420 & 1.565 & 1.506 \\
--1.4 & 1.262 & 1.222 & 1.335 & 1.283 & 1.406 & 1.336 & 1.601 & 1.448 & 1.660 & 1.544 & 1.715 & 1.640 \\
--1.3 & 1.394 & 1.353 & 1.478 & 1.423 & 1.558 & 1.484 & 1.771 & 1.592 & 1.833 & 1.689 & 1.890 & 1.798 \\
--1.2 & 1.548 & 1.503 & 1.642 & 1.580 & 1.731 & 1.650 & 1.964 & 1.754 & 2.029 & 1.839 & 2.089 & 1.960 \\
--1.1 & 1.720 & 1.663 & 1.824 & 1.747 & 1.923 & 1.824 & 2.177 & 1.930 & 2.246 & 1.995 & 2.309 & 2.129 \\
--1.0 & 1.906 & 1.833 & 2.021 & 1.922 & 2.128 & 2.007 & 2.405 & 2.121 & 2.480 & 2.168 & 2.547 & 2.295 \\
--0.9 & 2.103 & 2.030 & 2.227 & 2.127 & 2.344 & 2.218 & 2.644 & 2.367 & 2.726 & 2.395 & 2.799 & 2.483 \\
--0.8 & 2.306 & 2.238 & 2.440 & 2.342 & 2.566 & 2.439 & 2.890 & 2.617 & 2.980 & 2.639 & 3.060 & 2.681 \\
--0.7 & 2.513 & 2.450 & 2.655 & 2.561 & 2.790 & 2.666 & 3.140 & 2.880 & 3.237 & 2.902 & 3.326 & 2.915 \\
--0.6 & 2.721 & 2.667 & 2.871 & 2.787 & 3.014 & 2.899 & 3.390 & 3.150 & 3.495 & 3.180 & 3.592 & 3.177 \\
--0.5 & 2.929 & 2.883 & 3.088 & 3.015 & 3.238 & 3.136 & 3.637 & 3.418 & 3.751 & 3.469 & 3.856 & 3.469 \\
--0.4 & 3.139 & 3.103 & 3.305 & 3.243 & 3.462 & 3.371 & 3.882 & 3.679 & 4.003 & 3.732 & 4.116 & 3.770 \\
--0.3 & 3.353 & 3.326 & 3.525 & 3.473 & 3.688 & 3.608 & 4.124 & 3.931 & 4.251 & 4.003 & 4.370 & 4.050 \\
--0.2 & 3.572 & 3.553 & 3.751 & 3.705 & 3.918 & 3.845 & 4.364 & 4.185 & 4.496 & 4.271 & 4.619 & 4.331 \\
--0.1 & 3.802 & 3.784 & 3.986 & 3.942 & 4.156 & 4.081 & 4.605 & 4.438 & 4.737 & 4.528 & 4.862 & 4.602 \\
0.0 & 4.044 & 4.024 & 4.232 & 4.184 & 4.403 & 4.322 & 4.846 & 4.689 & 4.976 & 4.783 & 5.099 & 4.865 \\
0.1 & 4.298 & 4.270 & 4.489 & 4.433 & 4.660 & 4.574 & 5.088 & 4.940 & 5.212 & 5.034 & 5.329 & 5.115 \\
0.2 & 4.559 & 4.523 & 4.752 & 4.690 & 4.919 & 4.829 & 5.326 & 5.185 & 5.441 & 5.274 & 5.550 & 5.346 \\
0.3 & 4.815 & 4.779 & 5.008 & 4.948 & 5.170 & 5.086 & 5.550 & 5.415 & 5.655 & 5.495 & 5.754 & 5.564 \\
0.4 & 5.046 & 5.010 & 5.234 & 5.178 & 5.390 & 5.312 & 5.743 & 5.613 & 5.838 & 5.683 & 5.927 & 5.749 \\
0.5 & 5.214 & 5.188 & 5.394 & 5.347 & 5.543 & 5.474 & 5.878 & 5.753 & 5.967 & 5.819 & 6.051 & 5.875 \\
\hline
\end{tabular}
\vspace{0.3cm}
\tablecomments{The entire data of various colors and absorption indices are available at  http://web.yonsei.ac.kr/cosmic/data/YEPS.htm.}.
\end{center}
\end{table}

\clearpage
\begin{table}
\begin{center}
\scriptsize
\caption{$\langle$Fe$\rangle$ of the YEPS model GCs without ($w/o$) and with ($w$) HBs for [$\alpha$/Fe] = 0.14 and ages ($t$) of 7, 8, 9, 12, 13, and 14 Gyr. The models are computed using J10 fitting functions. \label{tbl-2}}
\begin{tabular}{ccccccccccccc}
\hline
[Fe/H] &
\multicolumn{12}{c}{$\langle$Fe$\rangle$} \\
\hline
{} &
\multicolumn{2}{c}{$t$ = 7} &
\multicolumn{2}{c}{8} &
\multicolumn{2}{c}{9} &
\multicolumn{2}{c}{12} &
\multicolumn{2}{c}{13} &
\multicolumn{2}{c}{14} \\
\hline
& $w/o$ & $w$ & $w/o$ & $w$ & $w/o$ & $w$ & $w/o$ & $w$ & $w/o$ & $w$ & $w/o$ & $w$ \\
\hline
--2.5 & 0.498 & 0.462 & 0.495 & 0.460 & 0.493 & 0.459 & 0.495 & 0.477 & 0.500 & 0.487 & 0.509 & 0.498 \\
--2.4 & 0.516 & 0.481 & 0.517 & 0.480 & 0.518 & 0.477 & 0.527 & 0.507 & 0.535 & 0.519 & 0.545 & 0.533 \\
--2.3 & 0.538 & 0.504 & 0.542 & 0.505 & 0.546 & 0.501 & 0.562 & 0.538 & 0.571 & 0.553 & 0.581 & 0.568 \\
--2.2 & 0.564 & 0.531 & 0.572 & 0.535 & 0.579 & 0.531 & 0.603 & 0.574 & 0.613 & 0.591 & 0.624 & 0.608 \\
--2.1 & 0.597 & 0.566 & 0.609 & 0.572 & 0.620 & 0.571 & 0.652 & 0.615 & 0.663 & 0.638 & 0.676 & 0.657 \\
--2.0 & 0.637 & 0.610 & 0.653 & 0.619 & 0.667 & 0.620 & 0.709 & 0.664 & 0.723 & 0.693 & 0.738 & 0.716 \\
--1.9 & 0.686 & 0.661 & 0.705 & 0.675 & 0.723 & 0.681 & 0.775 & 0.721 & 0.793 & 0.757 & 0.811 & 0.786 \\
--1.8 & 0.742 & 0.719 & 0.765 & 0.737 & 0.787 & 0.748 & 0.851 & 0.784 & 0.872 & 0.829 & 0.894 & 0.864 \\
--1.7 & 0.806 & 0.785 & 0.834 & 0.808 & 0.860 & 0.824 & 0.934 & 0.855 & 0.959 & 0.909 & 0.984 & 0.950 \\
--1.6 & 0.877 & 0.858 & 0.909 & 0.884 & 0.940 & 0.904 & 1.025 & 0.931 & 1.053 & 0.993 & 1.080 & 1.043 \\
--1.5 & 0.955 & 0.937 & 0.992 & 0.967 & 1.027 & 0.991 & 1.122 & 1.015 & 1.152 & 1.082 & 1.181 & 1.140 \\
--1.4 & 1.038 & 1.022 & 1.080 & 1.056 & 1.120 & 1.084 & 1.224 & 1.103 & 1.255 & 1.167 & 1.285 & 1.233 \\
--1.3 & 1.127 & 1.112 & 1.174 & 1.151 & 1.218 & 1.183 & 1.330 & 1.196 & 1.362 & 1.252 & 1.391 & 1.327 \\
--1.2 & 1.219 & 1.208 & 1.271 & 1.249 & 1.319 & 1.286 & 1.438 & 1.301 & 1.470 & 1.333 & 1.498 & 1.411 \\
--1.1 & 1.316 & 1.307 & 1.372 & 1.352 & 1.423 & 1.393 & 1.548 & 1.416 & 1.580 & 1.422 & 1.607 & 1.493 \\
--1.0 & 1.415 & 1.410 & 1.475 & 1.457 & 1.530 & 1.502 & 1.660 & 1.539 & 1.691 & 1.529 & 1.716 & 1.575 \\
--0.9 & 1.518 & 1.517 & 1.580 & 1.567 & 1.637 & 1.614 & 1.772 & 1.670 & 1.804 & 1.651 & 1.829 & 1.660 \\
--0.8 & 1.623 & 1.630 & 1.687 & 1.681 & 1.747 & 1.729 & 1.886 & 1.799 & 1.919 & 1.782 & 1.944 & 1.761 \\
--0.7 & 1.732 & 1.747 & 1.797 & 1.798 & 1.858 & 1.847 & 2.002 & 1.934 & 2.036 & 1.921 & 2.064 & 1.887 \\
--0.6 & 1.845 & 1.871 & 1.910 & 1.922 & 1.971 & 1.971 & 2.120 & 2.070 & 2.158 & 2.071 & 2.189 & 2.035 \\
--0.5 & 1.962 & 2.000 & 2.026 & 2.053 & 2.088 & 2.101 & 2.243 & 2.210 & 2.284 & 2.225 & 2.320 & 2.209 \\
--0.4 & 2.085 & 2.139 & 2.148 & 2.189 & 2.209 & 2.238 & 2.371 & 2.355 & 2.417 & 2.379 & 2.460 & 2.387 \\
--0.3 & 2.214 & 2.284 & 2.277 & 2.335 & 2.337 & 2.383 & 2.505 & 2.510 & 2.557 & 2.538 & 2.607 & 2.558 \\
--0.2 & 2.351 & 2.439 & 2.413 & 2.489 & 2.473 & 2.537 & 2.647 & 2.670 & 2.705 & 2.706 & 2.763 & 2.738 \\
--0.1 & 2.496 & 2.601 & 2.558 & 2.649 & 2.618 & 2.702 & 2.799 & 2.839 & 2.862 & 2.885 & 2.926 & 2.926 \\
0.0 & 2.649 & 2.768 & 2.713 & 2.818 & 2.773 & 2.874 & 2.960 & 3.016 & 3.027 & 3.067 & 3.097 & 3.117 \\
0.1 & 2.811 & 2.944 & 2.877 & 2.994 & 2.939 & 3.050 & 3.130 & 3.198 & 3.200 & 3.253 & 3.272 & 3.308 \\
0.2 & 2.981 & 3.121 & 3.049 & 3.174 & 3.114 & 3.233 & 3.308 & 3.383 & 3.377 & 3.439 & 3.450 & 3.497 \\
0.3 & 3.155 & 3.305 & 3.228 & 3.361 & 3.295 & 3.417 & 3.491 & 3.572 & 3.558 & 3.627 & 3.627 & 3.683 \\
0.4 & 3.330 & 3.483 & 3.407 & 3.547 & 3.478 & 3.599 & 3.672 & 3.759 & 3.736 & 3.813 & 3.799 & 3.864 \\
0.5 & 3.500 & 3.655 & 3.581 & 3.718 & 3.653 & 3.777 & 3.846 & 3.939 & 3.905 & 3.988 & 3.962 & 4.038 \\
\hline
\end{tabular}
\vspace{0.3cm}
\tablecomments{The entire data of various colors and absorption indices are available at  http://web.yonsei.ac.kr/cosmic/data/YEPS.htm.}.
\end{center}
\end{table}

\clearpage
\begin{table}
\begin{center}
\scriptsize
\caption{The H$\beta$ of the YEPS model GCs without ($w/o$) and with ($w$) HBs for [$\alpha$/Fe] = 0.14 and ages ($t$) of 7, 8, 9, 12, 13, and 14 Gyr. The models are computed using J10 fitting functions. \label{tbl-3}}
\begin{tabular}{ccccccccccccc}
\hline
[Fe/H] &
\multicolumn{12}{c}{H$\beta$} \\
\hline
{} &
\multicolumn{2}{c}{$t$ = 7} &
\multicolumn{2}{c}{8} &
\multicolumn{2}{c}{9} &
\multicolumn{2}{c}{12} &
\multicolumn{2}{c}{13} &
\multicolumn{2}{c}{14} \\
\hline
& $w/o$ & $w$ & $w/o$ & $w$ & $w/o$ & $w$ & $w/o$ & $w$ & $w/o$ & $w$ & $w/o$ & $w$ \\
\hline
--2.5 & 3.336 & 3.368 & 3.052 & 3.207 & 2.810 & 3.211 & 2.287 & 2.613 & 2.169 & 2.365 & 2.075 & 2.202 \\
--2.4 & 3.299 & 3.292 & 3.019 & 3.126 & 2.780 & 3.128 & 2.270 & 2.624 & 2.154 & 2.374 & 2.062 & 2.196 \\
--2.3 & 3.264 & 3.221 & 2.987 & 3.049 & 2.753 & 3.039 & 2.253 & 2.647 & 2.140 & 2.375 & 2.048 & 2.195 \\
--2.2 & 3.226 & 3.167 & 2.954 & 2.986 & 2.725 & 2.950 & 2.236 & 2.673 & 2.124 & 2.388 & 2.033 & 2.195 \\
--2.1 & 3.182 & 3.109 & 2.916 & 2.923 & 2.692 & 2.854 & 2.215 & 2.698 & 2.105 & 2.395 & 2.015 & 2.189 \\
--2.0 & 3.132 & 3.045 & 2.871 & 2.854 & 2.654 & 2.754 & 2.191 & 2.722 & 2.083 & 2.408 & 1.993 & 2.184 \\
--1.9 & 3.074 & 2.975 & 2.820 & 2.780 & 2.610 & 2.656 & 2.162 & 2.737 & 2.057 & 2.409 & 1.968 & 2.169 \\
--1.8 & 3.010 & 2.906 & 2.763 & 2.712 & 2.560 & 2.579 & 2.130 & 2.746 & 2.028 & 2.410 & 1.940 & 2.154 \\
--1.7 & 2.941 & 2.831 & 2.701 & 2.640 & 2.506 & 2.500 & 2.094 & 2.738 & 1.996 & 2.407 & 1.910 & 2.136 \\
--1.6 & 2.867 & 2.760 & 2.636 & 2.575 & 2.448 & 2.437 & 2.055 & 2.725 & 1.961 & 2.409 & 1.878 & 2.109 \\
--1.5 & 2.792 & 2.688 & 2.569 & 2.508 & 2.389 & 2.374 & 2.014 & 2.689 & 1.924 & 2.400 & 1.844 & 2.076 \\
--1.4 & 2.715 & 2.612 & 2.501 & 2.440 & 2.328 & 2.309 & 1.971 & 2.629 & 1.886 & 2.446 & 1.810 & 2.091 \\
--1.3 & 2.639 & 2.538 & 2.434 & 2.374 & 2.269 & 2.246 & 1.927 & 2.545 & 1.846 & 2.484 & 1.775 & 2.106 \\
--1.2 & 2.566 & 2.466 & 2.369 & 2.308 & 2.210 & 2.183 & 1.883 & 2.406 & 1.806 & 2.527 & 1.739 & 2.187 \\
--1.1 & 2.495 & 2.398 & 2.307 & 2.245 & 2.154 & 2.124 & 1.838 & 2.243 & 1.765 & 2.491 & 1.703 & 2.266 \\
--1.0 & 2.427 & 2.334 & 2.248 & 2.186 & 2.101 & 2.067 & 1.795 & 2.080 & 1.725 & 2.357 & 1.666 & 2.328 \\
--0.9 & 2.363 & 2.271 & 2.192 & 2.129 & 2.051 & 2.014 & 1.752 & 1.931 & 1.684 & 2.175 & 1.628 & 2.351 \\
--0.8 & 2.302 & 2.209 & 2.140 & 2.075 & 2.003 & 1.962 & 1.710 & 1.824 & 1.643 & 1.990 & 1.590 & 2.285 \\
--0.7 & 2.244 & 2.150 & 2.089 & 2.022 & 1.957 & 1.914 & 1.669 & 1.732 & 1.603 & 1.830 & 1.550 & 2.136 \\
--0.6 & 2.187 & 2.093 & 2.041 & 1.973 & 1.913 & 1.865 & 1.629 & 1.662 & 1.563 & 1.683 & 1.510 & 1.908 \\
--0.5 & 2.132 & 2.039 & 1.992 & 1.924 & 1.870 & 1.820 & 1.590 & 1.606 & 1.523 & 1.584 & 1.469 & 1.679 \\
--0.4 & 2.076 & 1.984 & 1.944 & 1.875 & 1.826 & 1.776 & 1.551 & 1.559 & 1.484 & 1.523 & 1.427 & 1.530 \\
--0.3 & 2.018 & 1.933 & 1.894 & 1.827 & 1.782 & 1.734 & 1.514 & 1.518 & 1.445 & 1.472 & 1.386 & 1.448 \\
--0.2 & 1.959 & 1.878 & 1.841 & 1.780 & 1.736 & 1.691 & 1.477 & 1.480 & 1.407 & 1.428 & 1.346 & 1.390 \\
--0.1 & 1.897 & 1.820 & 1.787 & 1.729 & 1.688 & 1.645 & 1.441 & 1.443 & 1.371 & 1.388 & 1.308 & 1.343 \\
0.0 & 1.834 & 1.761 & 1.731 & 1.677 & 1.639 & 1.598 & 1.406 & 1.405 & 1.337 & 1.351 & 1.272 & 1.302 \\
0.1 & 1.771 & 1.697 & 1.674 & 1.619 & 1.590 & 1.544 & 1.373 & 1.363 & 1.305 & 1.310 & 1.239 & 1.260 \\
0.2 & 1.711 & 1.633 & 1.620 & 1.559 & 1.543 & 1.489 & 1.341 & 1.323 & 1.276 & 1.272 & 1.210 & 1.221 \\
0.3 & 1.658 & 1.573 & 1.573 & 1.503 & 1.501 & 1.441 & 1.313 & 1.288 & 1.250 & 1.237 & 1.184 & 1.186 \\
0.4 & 1.619 & 1.529 & 1.538 & 1.460 & 1.470 & 1.407 & 1.290 & 1.258 & 1.228 & 1.208 & 1.162 & 1.155 \\
0.5 & 1.602 & 1.507 & 1.522 & 1.442 & 1.455 & 1.386 & 1.273 & 1.234 & 1.209 & 1.182 & 1.143 & 1.127 \\
\hline
\end{tabular}
\vspace{0.3cm}
\tablecomments{The entire data of various colors and absorption indices are available at  http://web.yonsei.ac.kr/cosmic/data/YEPS.htm.}.
\end{center}
\end{table}

\clearpage
\begin{table}
\begin{center}
\scriptsize
\caption{The H$\gamma_{F}$ of the YEPS model GCs without ($w/o$) and with ($w$) HBs for [$\alpha$/Fe] = 0.14 and ages ($t$) of 7, 8, 9, 12, 13, and 14 Gyr. The models are computed using J10 fitting functions. \label{tbl-4}}
\begin{tabular}{ccccccccccccc}
\hline
[Fe/H] &
\multicolumn{12}{c}{H$\gamma_{F}$} \\
\hline
{} &
\multicolumn{2}{c}{$t$ = 7} &
\multicolumn{2}{c}{8} &
\multicolumn{2}{c}{9} &
\multicolumn{2}{c}{12} &
\multicolumn{2}{c}{13} &
\multicolumn{2}{c}{14} \\
\hline
& $w/o$ & $w$ & $w/o$ & $w$ & $w/o$ & $w$ & $w/o$ & $w$ & $w/o$ & $w$ & $w/o$ & $w$ \\
\hline
--2.5 & 4.035 & 4.137 & 3.586 & 3.897 & 3.193 & 3.901 & 2.307 & 2.883 & 2.092 & 2.454 & 1.912 & 2.154 \\
--2.4 & 3.924 & 3.957 & 3.475 & 3.709 & 3.083 & 3.721 & 2.202 & 2.841 & 1.988 & 2.405 & 1.809 & 2.073 \\
--2.3 & 3.821 & 3.784 & 3.369 & 3.521 & 2.976 & 3.523 & 2.097 & 2.822 & 1.886 & 2.341 & 1.709 & 2.005 \\
--2.2 & 3.712 & 3.639 & 3.259 & 3.355 & 2.865 & 3.320 & 1.988 & 2.810 & 1.780 & 2.304 & 1.606 & 1.940 \\
--2.1 & 3.589 & 3.485 & 3.136 & 3.179 & 2.744 & 3.089 & 1.872 & 2.798 & 1.666 & 2.251 & 1.493 & 1.862 \\
--2.0 & 3.449 & 3.314 & 2.998 & 2.989 & 2.609 & 2.837 & 1.747 & 2.781 & 1.542 & 2.213 & 1.370 & 1.784 \\
--1.9 & 3.291 & 3.122 & 2.844 & 2.785 & 2.461 & 2.580 & 1.613 & 2.751 & 1.410 & 2.151 & 1.237 & 1.684 \\
--1.8 & 3.116 & 2.930 & 2.676 & 2.587 & 2.300 & 2.358 & 1.470 & 2.711 & 1.268 & 2.091 & 1.095 & 1.584 \\
--1.7 & 2.928 & 2.723 & 2.495 & 2.378 & 2.128 & 2.123 & 1.318 & 2.638 & 1.119 & 2.024 & 0.946 & 1.477 \\
--1.6 & 2.730 & 2.521 & 2.304 & 2.179 & 1.947 & 1.924 & 1.158 & 2.558 & 0.963 & 1.971 & 0.792 & 1.353 \\
--1.5 & 2.524 & 2.314 & 2.107 & 1.975 & 1.758 & 1.718 & 0.993 & 2.436 & 0.803 & 1.895 & 0.636 & 1.219 \\
--1.4 & 2.314 & 2.098 & 1.905 & 1.764 & 1.564 & 1.507 & 0.822 & 2.266 & 0.638 & 1.938 & 0.477 & 1.199 \\
--1.3 & 2.101 & 1.881 & 1.701 & 1.553 & 1.368 & 1.294 & 0.647 & 2.045 & 0.471 & 1.964 & 0.317 & 1.177 \\
--1.2 & 1.888 & 1.662 & 1.495 & 1.336 & 1.169 & 1.079 & 0.469 & 1.705 & 0.301 & 1.999 & 0.156 & 1.314 \\
--1.1 & 1.673 & 1.444 & 1.289 & 1.121 & 0.970 & 0.864 & 0.290 & 1.283 & 0.130 & 1.869 & --0.006 & 1.442 \\
--1.0 & 1.457 & 1.229 & 1.083 & 0.906 & 0.771 & 0.652 & 0.108 & 0.826 & --0.044 & 1.519 & --0.170 & 1.529 \\
--0.9 & 1.240 & 1.007 & 0.876 & 0.691 & 0.571 & 0.437 & --0.074 & 0.366 & --0.219 & 1.035 & --0.336 & 1.526 \\
--0.8 & 1.019 & 0.778 & 0.668 & 0.471 & 0.372 & 0.221 & --0.257 & --0.006 & --0.396 & 0.508 & --0.507 & 1.312 \\
--0.7 & 0.794 & 0.545 & 0.458 & 0.250 & 0.171 & 0.007 & --0.441 & --0.350 & --0.576 & --0.004 & --0.682 & 0.877 \\
--0.6 & 0.564 & 0.307 & 0.244 & 0.030 & --0.031 & --0.211 & --0.624 & --0.627 & --0.757 & --0.505 & --0.861 & 0.187 \\
--0.5 & 0.327 & 0.067 & 0.025 & --0.195 & --0.235 & --0.424 & --0.808 & --0.866 & --0.940 & --0.879 & --1.045 & --0.539 \\
--0.4 & 0.083 & --0.179 & --0.198 & --0.420 & --0.441 & --0.635 & --0.992 & --1.081 & --1.124 & --1.149 & --1.232 & --1.072 \\
--0.3 & --0.166 & --0.416 & --0.425 & --0.643 & --0.651 & --0.839 & --1.175 & --1.288 & --1.308 & --1.375 & --1.422 & --1.404 \\
--0.2 & --0.420 & --0.658 & --0.656 & --0.862 & --0.862 & --1.045 & --1.358 & --1.477 & --1.491 & --1.581 & --1.611 & --1.659 \\
--0.1 & --0.675 & --0.900 & --0.889 & --1.081 & --1.076 & --1.254 & --1.539 & --1.656 & --1.672 & --1.771 & --1.797 & --1.872 \\
0.0 & --0.928 & --1.135 & --1.122 & --1.301 & --1.289 & --1.460 & --1.717 & --1.827 & --1.849 & --1.944 & --1.978 & --2.055 \\
0.1 & --1.174 & --1.366 & --1.350 & --1.515 & --1.500 & --1.655 & --1.893 & --1.990 & --2.020 & --2.104 & --2.150 & --2.221 \\
0.2 & --1.407 & --1.579 & --1.569 & --1.716 & --1.705 & --1.844 & --2.065 & --2.148 & --2.185 & --2.257 & --2.311 & --2.373 \\
0.3 & --1.622 & --1.779 & --1.775 & --1.907 & --1.901 & --2.019 & --2.231 & --2.303 & --2.343 & --2.404 & --2.461 & --2.512 \\
0.4 & --1.812 & --1.949 & --1.961 & --2.079 & --2.082 & --2.181 & --2.391 & --2.452 & --2.493 & --2.545 & --2.600 & --2.642 \\
0.5 & --1.974 & --2.096 & --2.121 & --2.224 & --2.242 & --2.330 & --2.543 & --2.597 & --2.638 & --2.683 & --2.732 & --2.768 \\
\hline
\end{tabular}
\vspace{0.3cm}
\tablecomments{The entire data of various colors and absorption indices are available at  http://web.yonsei.ac.kr/cosmic/data/YEPS.htm.}.
\end{center}
\end{table}

\clearpage
\begin{table}
\begin{center}
\scriptsize
\caption{The H$\delta_{F}$ of the YEPS model GCs without ($w/o$) and with ($w$) HBs for [$\alpha$/Fe] = 0.14 and ages ($t$) of 7, 8, 9, 12, 13, and 14 Gyr. The models are computed using J10 fitting functions. \label{tbl-5}}
\begin{tabular}{ccccccccccccc}
\hline
[Fe/H] &
\multicolumn{12}{c}{H$\delta_{F}$} \\
\hline
{} &
\multicolumn{2}{c}{$t$ = 7} &
\multicolumn{2}{c}{8} &
\multicolumn{2}{c}{9} &
\multicolumn{2}{c}{12} &
\multicolumn{2}{c}{13} &
\multicolumn{2}{c}{14} \\
\hline
& $w/o$ & $w$ & $w/o$ & $w$ & $w/o$ & $w$ & $w/o$ & $w$ & $w/o$ & $w$ & $w/o$ & $w$ \\
\hline
--2.5 & 4.163 & 4.142 & 3.715 & 3.875 & 3.337 & 3.864 & 2.529 & 3.157 & 2.342 & 2.746 & 2.186 & 2.459 \\
--2.4 & 4.082 & 4.021 & 3.643 & 3.751 & 3.275 & 3.736 & 2.492 & 3.175 & 2.310 & 2.766 & 2.157 & 2.447 \\
--2.3 & 4.001 & 3.898 & 3.570 & 3.622 & 3.211 & 3.594 & 2.451 & 3.210 & 2.275 & 2.761 & 2.127 & 2.444 \\
--2.2 & 3.910 & 3.791 & 3.488 & 3.505 & 3.137 & 3.444 & 2.399 & 3.241 & 2.230 & 2.779 & 2.088 & 2.440 \\
--2.1 & 3.805 & 3.669 & 3.392 & 3.377 & 3.051 & 3.273 & 2.336 & 3.266 & 2.172 & 2.776 & 2.035 & 2.417 \\
--2.0 & 3.684 & 3.531 & 3.283 & 3.233 & 2.952 & 3.086 & 2.261 & 3.278 & 2.102 & 2.781 & 1.968 & 2.388 \\
--1.9 & 3.548 & 3.378 & 3.160 & 3.076 & 2.841 & 2.893 & 2.175 & 3.272 & 2.020 & 2.760 & 1.889 & 2.336 \\
--1.8 & 3.400 & 3.222 & 3.025 & 2.925 & 2.719 & 2.727 & 2.080 & 3.247 & 1.930 & 2.737 & 1.800 & 2.281 \\
--1.7 & 3.242 & 3.054 & 2.882 & 2.764 & 2.589 & 2.554 & 1.978 & 3.192 & 1.832 & 2.707 & 1.705 & 2.220 \\
--1.6 & 3.077 & 2.890 & 2.732 & 2.611 & 2.453 & 2.406 & 1.871 & 3.121 & 1.731 & 2.688 & 1.608 & 2.144 \\
--1.5 & 2.910 & 2.722 & 2.580 & 2.453 & 2.314 & 2.254 & 1.761 & 3.015 & 1.628 & 2.649 & 1.511 & 2.062 \\
--1.4 & 2.743 & 2.550 & 2.427 & 2.293 & 2.173 & 2.098 & 1.650 & 2.855 & 1.525 & 2.717 & 1.416 & 2.086 \\
--1.3 & 2.578 & 2.380 & 2.276 & 2.135 & 2.034 & 1.945 & 1.540 & 2.649 & 1.424 & 2.756 & 1.324 & 2.109 \\
--1.2 & 2.415 & 2.214 & 2.128 & 1.976 & 1.898 & 1.793 & 1.432 & 2.344 & 1.326 & 2.773 & 1.235 & 2.269 \\
--1.1 & 2.257 & 2.051 & 1.983 & 1.822 & 1.764 & 1.645 & 1.325 & 1.999 & 1.229 & 2.632 & 1.149 & 2.413 \\
--1.0 & 2.103 & 1.896 & 1.843 & 1.672 & 1.635 & 1.504 & 1.221 & 1.656 & 1.134 & 2.310 & 1.064 & 2.499 \\
--0.9 & 1.952 & 1.740 & 1.707 & 1.526 & 1.509 & 1.361 & 1.119 & 1.342 & 1.039 & 1.914 & 0.977 & 2.494 \\
--0.8 & 1.803 & 1.585 & 1.574 & 1.383 & 1.387 & 1.224 & 1.018 & 1.100 & 0.944 & 1.511 & 0.888 & 2.284 \\
--0.7 & 1.656 & 1.434 & 1.443 & 1.243 & 1.268 & 1.093 & 0.917 & 0.884 & 0.846 & 1.152 & 0.794 & 1.927 \\
--0.6 & 1.510 & 1.285 & 1.313 & 1.113 & 1.150 & 0.964 & 0.816 & 0.719 & 0.746 & 0.809 & 0.694 & 1.361 \\
--0.5 & 1.363 & 1.139 & 1.184 & 0.983 & 1.034 & 0.845 & 0.713 & 0.581 & 0.642 & 0.573 & 0.586 & 0.817 \\
--0.4 & 1.215 & 0.996 & 1.054 & 0.856 & 0.916 & 0.730 & 0.608 & 0.459 & 0.534 & 0.402 & 0.471 & 0.451 \\
--0.3 & 1.067 & 0.862 & 0.923 & 0.734 & 0.798 & 0.622 & 0.501 & 0.337 & 0.421 & 0.274 & 0.349 & 0.240 \\
--0.2 & 0.918 & 0.728 & 0.790 & 0.616 & 0.678 & 0.512 & 0.390 & 0.235 & 0.305 & 0.157 & 0.223 & 0.087 \\
--0.1 & 0.771 & 0.593 & 0.657 & 0.496 & 0.555 & 0.394 & 0.276 & 0.134 & 0.186 & 0.045 & 0.095 & --0.042 \\
0.0 & 0.627 & 0.462 & 0.524 & 0.372 & 0.431 & 0.274 & 0.159 & 0.032 & 0.066 & --0.061 & --0.031 & --0.156 \\
0.1 & 0.488 & 0.330 & 0.392 & 0.245 & 0.305 & 0.158 & 0.042 & --0.074 & --0.053 & --0.167 & --0.153 & --0.267 \\
0.2 & 0.356 & 0.208 & 0.264 & 0.127 & 0.179 & 0.041 & --0.076 & --0.182 & --0.169 & --0.272 & --0.267 & --0.373 \\
0.3 & 0.232 & 0.090 & 0.140 & 0.011 & 0.056 & --0.068 & --0.193 & --0.292 & --0.280 & --0.378 & --0.371 & --0.470 \\
0.4 & 0.117 & --0.013 & 0.022 & --0.099 & --0.064 & --0.175 & --0.305 & --0.400 & --0.385 & --0.481 & --0.466 & --0.560 \\
0.5 & 0.010 & --0.111 & --0.088 & --0.198 & --0.176 & --0.281 & --0.413 & --0.507 & --0.485 & --0.579 & --0.555 & --0.650 \\
\hline
\end{tabular}
\vspace{0.3cm}
\tablecomments{The entire data of various colors and absorption indices are available at  http://web.yonsei.ac.kr/cosmic/data/YEPS.htm.}.
\end{center}
\end{table}

\clearpage
\begin{table*}
\begin{center}
\scriptsize
\caption{The result of the GMM$^{a}$ analysis for the distributions shown in Figure 4. \label{tbl-6}}
\begin{tabular}{lcccc}
\hline
\hline
& Mg{\it b} & H$\beta$ & H$\gamma_{F}$ &  H$\delta_{F}$ \\
\hline
  &  \multicolumn{4}{c}{{\it P}-value$^{b}$} \\
  & $p(\chi^2)$ $p(DD)$ $p(kurt)$ & $p(\chi^2)$ $p(DD)$ $p(kurt)$& $p(\chi^2)$ $p(DD)$ $p(kurt)$& $p(\chi^2)$ $p(DD)$ $p(kurt)$ \\
  \hline
Observations								& 0.001 0.374 0.636 & 0.001 0.088 0.001 & 0.001 0.107 0.001 & 0.001 0.061 0.001 \\
Model$^{c}$ (J10)								& 0.010  0.010  0.010 & 0.010  0.010  0.010 & 0.010  0.010  0.010 & 0.010  0.010  0.010 \\
Model$^{d}$ (S07 \& W94)							& 0.010  0.010  0.010 & 0.010  0.010  0.010 & 0.010  0.010  0.010 & 0.010  0.010  0.010 \\
\hline		    
  &  \multicolumn{4}{c}{Number Ratio$^{e}$} \\	  
\hline
Observations								& 0.399\,:\,0.601 & 0.517\,:\,0.483 & 0.558\,:\,0.442 & 0.525\,:\,0.475 \\

Models$^{c}$ (J10)								& 0.406\,:\,0.594 & 0.621\,:\,0.379 & 0.653\,:\,0.347 & 0.674\,:\,0.326 \\

Models$^{d}$ (S07 \& W94)							& 0.358\,:\,0.642 & 0.551\,:\,0.449 & 0.638\,:\,0.362 & 0.622\,:\,0.378 \\

\hline
  &  $\mu^{f}_{mp}$, $\mu^{f}_{mr}$  & $\mu_{mp}$, $\mu_{mr}$ & $\mu_{mp}$, $\mu_{mr}$ & $\mu_{mp}$, $\mu_{mr}$\\ 
  & ($\sigma^{f}_{mp}$, $\sigma^{f}_{mr}$) & ($\sigma_{mp}$, $\sigma_{mr}$) & ($\sigma_{mp}$, $\sigma_{mr}$) & ($\sigma_{mp}$, $\sigma_{mr}$) \\
\hline
Observation								&  1.073, 2.679 & 1.675, 2.455 & --0.473, 1.824 & 0.747, 2.616 \\
										& (0.329, 1.031) & (0.212, 0.219) & (0.870, 0.451) & (0.532, 0.391) \\
Models$^{c}$ (J10)  			& 1.572, 3.308 & 1.694, 2.585 & -0.489, 2.281 & 0.868, 2.913 \\
										& (0.467, 1.111) & (0.294, 0.212) & (1.137, 0.550) & (0.754, 0.423) \\
Models$^{d}$ (S07 \& W94)	& 1.420, 2.921 & 1.618, 2.360 & -0.051, 2.050 & 0.935, 2.536 \\
										& (0.446, 0.910) & (0.233, 0.193) & (0.952, 0.489) & (0.633, 0.365) \\				
\hline
\end{tabular}
\vspace{0.3cm}
\tablecomments{
$^{a}$ GMM \citep{muratov10} is a statistical technique for detecting and quantifying bimodality in a distribution.\\
$^{b}$ The resulting significance of the GMM test is expressed in terms of {\it P}-value. $P\,<$ 0.05 indicates that bimodality of a distribution is deemed significant.\\ 
$^{c}$ The YEPS models computed with the fitting functions by J10.\\
$^{d}$ The YEPS models computed with the fitting functions by S07 and W94.\\
$^{e}$ The number ratios between index-weak and strong groups.\\ 
$^{f}$ The mean and the standard deviations of metal-poor (mp) and metal-rich (mr) groups.}. 
\end{center}
\end{table*}

\clearpage
\begin{table*}
\begin{center}
\scriptsize
\caption{The result of the GMM analysis for the distributions shown in Figure 5. \label{tbl-7}}
\begin{tabular}{lcccc}
\hline
\hline
& Mg{\it b} & H$\beta$ & H$\gamma_{F}$ &  H$\delta_{F}$ \\
\hline
  &  \multicolumn{4}{c}{{\it P}-value$^{a}$} \\
  & $p(\chi^2)$ $p(DD)$ $p(kurt)$ & $p(\chi^2)$ $p(DD)$ $p(kurt)$& $p(\chi^2)$ $p(DD)$ $p(kurt)$& $p(\chi^2)$ $p(DD)$ $p(kurt)$ \\
  \hline
Observations								& 0.001 0.374 0.636 & 0.001 0.088 0.001 & 0.001 0.107 0.001 & 0.001 0.061 0.001 \\
Conventional Linear Projection					& 0.115 0.053 0.762 & 0.824 0.301 0.374 & 0.772 0.245 0.249 & 0.756 0.246 0.236 \\
Models$^{b}$ (Nonlinear Projection, J10) 			& 0.001 0.576 0.841 & 0.001 0.109 0.001 & 0.001 0.113 0.001 & 0.001 0.108 0.001 \\
Models$^{c}$ (Nonlinear Projection, S07 \& W94) 	& 0.001 0.509 0.067 & 0.001 0.096 0.001 & 0.001 0.122 0.001 & 0.001 0.115 0.001 \\
\hline		    
  &  \multicolumn{4}{c}{Number Ratio$^{d}$} \\	  
\hline
Observations								& 0.399\,:\,0.601 & 0.517\,:\,0.483 & 0.558\,:\,0.442 & 0.525\,:\,0.475 \\
Conventional Linear Projection					& 0.014\,:\,0.986 & 0.947\,:\,0.053 & 0.960\,:\,0.040 & 0.958\,:\,0.042 \\
Models$^{b}$ (Nonlinear Projection, J10)			& 0.302\,:\,0.698 & 0.453\,:\,0.547 & 0.479\,:\,0.521 & 0.473\,:\,0.527 \\
Models$^{c}$ (Nonlinear Projection, S07 \& W94)	& 0.274\,:\,0.726 & 0.452\,:\,0.548 & 0.476\,:\,0.524 & 0.512\,:\,0.488 \\
\hline
  &  $\mu^{e}_{mp}$, $\mu^{f}_{mr}$  & $\mu_{mp}$, $\mu_{mr}$ & $\mu_{mp}$, $\mu_{mr}$ & $\mu_{mp}$, $\mu_{mr}$\\ 
  & ($\sigma^{e}_{mp}$, $\sigma^{f}_{mr}$) & ($\sigma_{mp}$, $\sigma_{mr}$) & ($\sigma_{mp}$, $\sigma_{mr}$) & ($\sigma_{mp}$, $\sigma_{mr}$) \\
\hline
Observation								&  1.073, 2.679 & 1.675, 2.455 & --0.473, 1.824 & 0.747, 2.616 \\
										& (0.329, 1.031) & (0.212, 0.219) & (0.870, 0.451) & (0.532, 0.391) \\
Conventional Linear Projection					& --1.452, 2.115 & 2.011, 3.032 & 0.446, 3.208 & 1.555, 3.741 \\
										& (0.294, 1.177) & (0.529, 0.333) & (1.430, 0.655) & (1.128, 0.513) \\
Models$^{c}$ (Nonlinear Projection, J10)  			&  1.334, 2.675 & 1.673, 2.523 & --0.531, 2.118 & 0.776, 2.710 \\
										& (0.341, 1.114) & (0.241, 0.259) & (0.915, 0.675) & (0.578, 0.570) \\
Models$^{d}$ (Nonlinear Projection, S07 \& W94)	&  1.210, 2.729 & 1.545, 2.314 & --0.476, 1.877 & 0.745, 2.435 \\
										& (0.390, 1.133) & (0.203, 0.224) & (0.831, 0.627) & (0.594, 0.436) \\
\hline
\end{tabular}
\vspace{0.3cm}
\tablecomments{
$^{a}$ The resulting significance of the GMM test expressed as {\it P}-value. \\
$^{b}$ The YEPS models computed with the fitting functions by J10.\\
$^{c}$ The YEPS models computed with the fitting functions by S07 and W94.\\
$^{d}$ The number ratios between index-weak and strong groups.\\ 
$^{e}$ The mean and the standard deviations of metal-poor (mp) and metal-rich (mr) groups.}. 
\end{center}
\end{table*}

\clearpage
\begin{table*}
\begin{center}
\scriptsize
\caption{The result of the GMM analysis for the distributions shown in Figure 6. \label{tbl-8}}
\begin{tabular}{lcccc}
\hline
\hline
& Mg{\it b} & H$\beta$ & H$\gamma_{F}$ &  H$\delta_{F}$ \\
\hline
  &  \multicolumn{4}{c}{{\it P}-value$^{a}$} \\
  & $p(\chi^2)$ $p(DD)$ $p(kurt)$ & $p(\chi^2)$ $p(DD)$ $p(kurt)$& $p(\chi^2)$ $p(DD)$ $p(kurt)$& $p(\chi^2)$ $p(DD)$ $p(kurt)$ \\
  \hline
Observations$^{b}$							& 0.466  0.060  0.651 & 0.466  0.060  0.651 & 0.466  0.060  0.651 & 0.466  0.060  0.651 \\
Conventional Linear Projection					& 0.010  0.010  1.000 & 0.010  0.010  0.010 & 0.010  0.010  0.010 & 0.010  0.010  0.010 \\
Models$^{b}$ (Nonlinear Projection, J10) 		 	& 0.010  0.510  1.000 & 0.010  0.040  1.000 & 0.010  0.740  1.000 & 0.010  0.030  1.000 \\
Models$^{c}$ (Nonlinear Projection, S07 \& W94) 	& 0.010  0.540  1.000 & 0.010  0.030  1.000 & 0.010  0.050  1.000 & 0.010  0.030  1.000 \\
\hline		    
  &  \multicolumn{4}{c}{Number Ratio$^{e}$} \\	  
\hline
Observations$^{b}$						 	& 0.010\,:\,0.990 & 0.010\,:\,0.990 & 0.010\,:\,0.990 & 0.010\,:\,0.990 \\
Conventional Linear Projection				 	& 0.381\,:\,0.619 & 0.475\,:\,0.525 & 0.445\,:\,0.555 & 0.486\,:\,0.514 \\
Models$^{b}$ (Nonlinear Projection, J10) 		  	& 0.595\,:\,0.405 & 0.862\,:\,0.138 & 0.122\,:\,0.878 & 0.882\,:\,0.118 \\
Models$^{c}$ (Nonlinear Projection, S07 \& W94) 	& 0.625\,:\,0.375 & 0.703\,:\,0.297 & 0.657\,:\,0.343 & 0.884\,:\,0.116 \\
\hline
  &  $\mu^{f}_{mp}$, $\mu^{f}_{mr}$  & $\mu_{mp}$, $\mu_{mr}$ & $\mu_{mp}$, $\mu_{mr}$ & $\mu_{mp}$, $\mu_{mr}$\\ 
  & ($\sigma^{f}_{mp}$, $\sigma^{f}_{mr}$) & ($\sigma_{mp}$, $\sigma_{mr}$) & ($\sigma_{mp}$, $\sigma_{mr}$) & ($\sigma_{mp}$, $\sigma_{mr}$) \\
\hline
Observations$^{b}$							& --2.660, --1.036 & --2.660, --1.036 & --2.660, --1.036 & --2.660, --1.036 \\
										& (0.138, 0.553) & (0.138, 0.553) & (0.138, 0.553) & (0.138, 0.553) \\
Conventional Linear Projection					& -1.504, -0.775 & -1.491, -0.651 & -1.552, -0.650 & -1.534, -0.599 \\
										& (0.180,  0.509) & (0.263,  0.270) & (0.203,  0.354) & (0.224,  0.272) \\
Models$^{b}$ (Nonlinear Projection, J10) 		 	& --1.168, --0.878 & --0.940,  -0.302 & --1.058, --0.940 & --1.018, --0.224 \\
							 			& (0.745, 0.428) & (0.353, 1.138) & (1.121, 0.382) & (0.405, 0.749) \\
Models$^{c}$ (Nonlinear Projection, S07 \& W94) 	& --1.118, --0.886 & --0.964,  0.319 & --1.141, --0.593 & --0.867,  0.214 \\
										& (0.422, 0.745) & (0.287, 1.146) & (0.370, 0.517) & (0 .453, 0.897) \\
\hline
\end{tabular}
\vspace{0.3cm}
\tablecomments{
$^{a}$ The resulting significance of the GMM test expressed as {\it P}-value. \\
$^{b}$ The observed [Fe/H] distribution as shown along the y-axes in Figure 5.\\
$^{c}$ The YEPS models computed with the fitting functions by J10.\\
$^{d}$ The YEPS models computed with the fitting functions by S07 and W94.\\
$^{e}$ The number ratios between index-weak and strong groups.\\
$^{f}$ The mean and the standard deviations of metal-poor (mp) and metal-rich (mr) groups.}. 
\end{center}
\end{table*}

\end{document}